\documentclass[12pt,a4paper]{article}
\usepackage[english]{babel}
\usepackage{amsmath,amssymb,enumerate,fullpage,setspace}
\usepackage{amsfonts}
\usepackage{bm}
\usepackage[authoryear,longnamesfirst]{natbib}
\usepackage{rotating,graphicx}
\usepackage[table]{xcolor}
\usepackage{footmisc}
\usepackage[flushleft]{threeparttable}
\usepackage{float}
\usepackage{caption}
\usepackage{subcaption}
\begin{document}
\title{On the Relationship between Markov Switching Models and Fuzzy Clustering: a Nonparametric Method to Detect the Number of States}
\author{Edoardo Otranto\\Department of Economics, University of Messina, and CRENoS\\ email: eotranto@unime.it\\ \\
Luca Scaffidi Domianello\\Department of Statistics, Computer Science, Applications, University of Florence\\ email: luca.scaffididomianello@unifi.it}
\date{} 

\maketitle
\begin{abstract}
Markov Switching models have had increasing success in time series analysis due to their ability to capture the existence of unobserved discrete states in the dynamics of the variables under study. This result is generally obtained thanks to the inference on states derived from the so--called Hamilton filter. One of the open problems in this framework is the identification of the number of states, generally fixed a priori; it is in fact impossible to apply classical tests due to the problem of the nuisance parameters present only under the alternative hypothesis. In this work we show, by Monte Carlo simulations, that fuzzy clustering is able to reproduce the parametric state inference derived from the Hamilton filter and that the typical indices used in clustering to determine the number of groups can be used to identify the number of states in this framework. The procedure is very simple to apply, considering that it is performed (in a nonparametric way) independently of the data generation process and that the indicators we use are present in most statistical packages. A final application on real data completes the analysis.
\end{abstract}

{\bf Keywords:} Nuisance parameters, groups identification, Monte Carlo simulations, Markov chains

\newpage
\section{Introduction}\label{sec:intro}

Markov Switching (MS) models \citep{ham90} have received increasing attention in time series analysis with several applications in economics \citep[see, for example][]{Hamilton2017}, finance \citep{go15}, neuroscience \citep{Degrasetal2022}, just to name a few fields. Its main advantage consists in the possibility to consider the existence of several states, interpreted as particular regimes (for example, expansion and contraction in business cycle, quiet and turmoil periods in the financial markets), which are not observed, but whose dynamics can be represented by an ergodic Markov chain. Thanks to the properties of  ergodic Markov chains, it is possible to make inference on the unobserved state, assigning to each state at each time a filtered or smoothed probability obtained from the so--called Hamilton filter \citep{ham90}. An open problem is the identification of the number of states, which is not feasible with classical tests for the problem of nuisance parameters present only under the alternative hypothesis.  In practice, when the model under the null hypothesis is a MS model with $k$ states and the model under the alternative is a model with $k+1$ states, the two models are not nested due to the presence of transition probabilities referring to the state $k+1$, not identified under the null hypothesis \citep[see, for example,][]{Hansen92}.  As known, in this framework the classical tests, such as Likelihood Ratio (LR), do not follow the standard distributions and the true distribution is unknown. There are several proposals to bypass this problem; the first approaches were based on the supremum of a LR test \citep{Davies77}, trying to derive its asymptotic distribution over a range of nuisance parameters via simulations \citep{Hansen92}, but with a high computational cost. Alternatively, \cite{Garcia98} proposes a similar approach by reducing the range of nuisance parameters, with the exclusion of some important particular cases.  Most recent approaches avoid the development of tests, trying to directly identify  the number of states by penalized likelihood criteria \citep{Psadarakis_Spagnolo2003,Psadarakis_Spagnolo2006} or Kullback–Leibler divergence \citep{Smithetal2006}.  

A very practical idea is to identify the number of states before the estimation step, emphasizing  the fact that in an MS models each observation is generated by a mixture density, with a number of components equal to the number of states of the Markov chain. This approach was developed in a nonparametric Bayesian framework by \cite{Otranto_Gallo2002}, where the posterior distribution of the number of states is derived by means of a Gibbs sampler. This intuition is the basis of our proposal.

First, we note a similarity between the inference on the regime in MS models and the grouping derived from a fuzzy clustering \citep{Durso2015}. This last approach is able to provide a clustering of statistical units in $k$ groups, with a probability of belonging to each  of the $k$ clusters. 

The relationship between mixture models and fuzzy clustering has been underlined in several works. For example, \cite{Davenport1988} compare the constrained Maximum Likelihood (ML) estimator of the parameters of a Normal mixture density with another based on fuzzy c--means clustering; an interesting result is that the initialization of the algorithm to maximize the likelihood with fuzzy estimates improves the computation time and the accuracy of the estimates. \cite{Hathaway1986} explores the relationship between some clustering methods and the EM algorithm (also used for the estimation of MS models), by decomposing the likelihood function into a part depending on the hard k--means objective function and a penalty term for soft partitions; this idea was extended to the fuzzy case by \cite{Ichihashi2001}. Recently, \cite{Serafini2023} have compared model--based methods, derived from mixtures of Gaussian and $t$ densities, and fuzzy methods for soft clustering.

 Our first analysis consists in verifying whether the fuzzy approach provides a grouping similar to the one derived from the inference on regimes of the MS models, using the same assignment criterion (the unit is assigned to the group corresponding to the mode). This exercise is performed by means of Monte Carlo experiments. After verifying the similarity of the results in terms of state inference, we verify whether the typical indices used to select the number of clusters are able to identify the true number of states. The resulting approach is very simple, it is applied in the identification phase, before estimating the model, as in \cite{Otranto_Gallo2002}, but unlike this, it is very fast and uses tools implemented in the main statistical routines.

The structure of the paper is as follows: in the next section the main features of the MS model and of fuzzy clustering are briefly recalled; in Section \ref{sec:mc} we present a large set of Monte Carlo experiments to verify the ability of fuzzy clustering to reproduce the inference on the states of MS models (subsection \ref{sec:inf}) and   the performance of several indices in detecting the number of states (subsection \ref{sec:id}). Section \ref{sec:appl} contains an empirical analysis on a real dataset, referring to the Gross Domestic Product (GDP) series of U.S., for which we have a subjective classification of the two phases of business cycle provided by NBER. Some final remarks will conclude the paper.

\section{MS Models and Fuzzy Clustering}\label{sec:msfuz}
Let us consider a time series $y_t$ with $t=1,\dots,T$. We say that its Data Generating Process (DGP) follows an MS process with $k$ states if:
\begin{equation}
y_t=f(\bm{x};\bm{\theta}_{s_t})+\varepsilon_t \label{ms_gen}
\end{equation} 
where $\bm{x}$ is a vector of exogenous variables, possibly including lagged values of $y_t$,  $\varepsilon_t$ ($t=1,\dots,T$) are independent disturbances with zero mean and  constant variance. The vector of unknown parameters $\bm{\theta}_{s_t}$ depends on an unobservable discrete random variable $s_t$,\footnote {The variance could also depend on $s_t$; we will not consider this case in our experiments, but the generalization is obvious.} which can assume values $1, 2, \dots k$, and whose dynamics is driven by an ergodic Markov chain, with elements of the transition probability matrix:
\begin{equation}
p_{ij}=Pr(s_t=j|s_{t-1}=i), \qquad i,j=1,\dots,k    \label{tranprob}
\end{equation}
with $\sum_{j=1}^k p_{ij}=1$ for each $i=1,\dots,k$. Also the variance of  $\varepsilon_t$ could depend on the state variable $s_t$, so they could be not identically distributed. A typical model used in an MS framework is the $MS-AR(p)$ process:
\begin{equation}
y_t=\mu_{s_t}+\sum_{i=1}^p \phi_i \left(y_{t-i}-\mu_{s_{t-i}}\right)+\varepsilon_t, \qquad  \varepsilon_t \sim N(0,\sigma^2)  \label{ms_ar}
\end{equation}
Under stationarity constraints, specification (\ref{ms_ar}) involves a time--varying unconditional mean of the process,  $\mu_{s_t}$, depending on the state at time $t$. The transition probabilities $p_{ij}$ are estimated with other parameters $\bm{\theta}_{s_t}=\left(\mu_1,\dots,\mu_k\right)'$, $\bm{\phi}=\left(\phi_1,\dots,\phi_p\right)'$ and $\sigma^2$ by Maximum Likelihood Estimator (MLE), deriving the likelihood function by means of the so--called Hamilton filter \citep{ham90}. 
The same Hamilton filter provides the possibility to specify, at each time $t$, the conditional probability to fall in a certain state $j$; given the information set ${I}_{\tau}=\left(y_{\tau}, y_{\tau-1}, \dots\right)$, it is possible to obtain the so--called  filtered probabilities $Pr\left(s_t=j|I_{t}\right)$, the predicted probabilities $Pr\left(s_t=j|I_{t-1}\right)$ and the smoothed probabilities $Pr\left(s_t=j|I_{T}\right)$.  The predicted probabilities enter the likelihood function; in fact the conditional density of each $y_t$ is expressed as:
\begin{equation}
f(y_t|I_{t-1};\bm{\theta}_{s_t},\bm{\phi},\sigma^2)=\sum_{j=1}^k f(y_t|s_{t}=j,I_{t-1};\bm{\theta}_{s_t},\bm{\phi},\sigma^2) Pr\left(s_t=j|I_{t-1}\right) \label{density}
\end{equation}
which is a mixture of distributions with weights represented by the predicted probabilities.

Moreover, the Hamilton filter provides also an intuitive inference on the regime, assigning to state $j$ the observation at time $t$ with mode in $j$, using as mass distribution the filtered or, more frequently, the smoothed probabilities.

As said in Section \ref{sec:intro}, the identification of the number of states $k$ is not achievable through classical statistical tests for the problem of nuisance parameters present only under the alternative hypothesis. As an example, let us suppose to verify the null hypothesis of a linear model (no states or $k=1$):
\begin{equation}
y_t=\mu+\varepsilon_t, \qquad  \varepsilon_t \sim N(0,\sigma^2) \label{lin_mod}
\end{equation}
against the alternative of a MS model with $k=2$ states:
\begin{equation}
\begin{array}{l}
y_t=\mu_{s_t}+\varepsilon_t, \qquad  \varepsilon_t \sim N(0,\sigma^2), \qquad s_t=1,2\\
\\ \label{ms_2}
\bm{P}=\left[
\begin{array}{cc}
p_{11}&1-p_{11}\\
1-p_{22}&p_{22} \end{array}
\right]
\end{array}
\end{equation}
The linear model can not be obtained from the MS model simply imposing that the switching parameters are equal in the two states ($\mu_1=\mu_2$), because the transition probabilities $p_{11}$ and $p_{22}$ are not identified under the null hypothesis and the two models are not nested. Similarly if the 2--state model (\ref{ms_2}) is  assumed under the null hypothesis, whereas the alternative refers to a MS model with $k=3$:
\begin{equation}
\begin{array}{l}
y_t=\mu_{s_t}+\varepsilon_t, \qquad  \varepsilon \sim N(0,\sigma^2), \qquad s_t=1,2,3\\
\\ \label{ms_3}
\bm{P}=\left[
\begin{array}{ccc}
p_{11}&p_{12}&p_{13}\\
p_{21}&p_{22}&p_{23}\\
p_{31}&p_{32}&p_{33} \end{array}
\right]; \qquad \sum\limits_{j=1}^3 p_{ij}=1, \quad i=1,2,3
\end{array}
\end{equation}
 In this case 4 probabilities in $\bm{P}$  are not identified under the null hypothesis.

There is some similarity between the MS models and the fuzzy clustering methods. The latter detect the belonging of each statistical unit to a cluster with a certain probability (the {\it membership grade}), unlike the classical hard clustering, where each unit can belong to exactly one cluster only. Clustering based on the degree of membership in each group can be seen as similar to state inference done using the smoothed probabilities in an MS framework. For example, considering the most popular fuzzy clustering algorithm, the {\it fuzzy k--means} \citep{Bezdek81},  the fuzzy partition  in $k$ groups of the observed $y_t$'s ($t=1,\dots,T$) is obtained by  minimizing:
\begin{equation}
\min\limits_{\bm{U},\bm{c}} \sum_{t=1}^T \sum_{j=1}^k u_{tj}^m d^2\left(y_t,c_j\right) \label{f_cen}
\end{equation}
where $\bm{U}=\left\{u_{t,j}\right\}$ ($t=1,\dots,T$; $j=1,\dots,k$) is the membership grade matrix, $\bm{c}=\left(c_1,\dots,c_k\right)'$ is the vector of centroids, $m$ is the fuzziness parameter, which tunes the degree of fuzziness, $d(\cdot,\cdot)$ is a distance measure. 

Each column of $\bm{U}$ can be interpreted similarly to the smoothed probabilities of an MS model. Based on this insight, we ask whether nonparametric methods used to detect the number of groups in clustering can be used as a method to identify the number of states in an MS model. The detection of the number of clusters is generally carried out by means of indicators implemented in the main statistical packages, making this approach very simple and easily usable even by non--experts.

The verification of the previous idea can be performed in two steps through Monte Carlo experiments where data are generated from several MS processes: first, fixing the {\it true} $k$, we compare the smoothed probabilities derived from the estimated MS model with the corresponding grade of membership matrix derived from (\ref{f_cen});\footnote{Classifying the European Central Bank announcements, \cite{Gallo_Lacava_Otranto_2021} find very similar results between the classification derived from smoothed probabilities of their MS model and the {\it k--means} clustering procedure.}  then we use the main indices to detect the number of clusters verifying if they are able to identify the correct number of states of the MS DGP. More specifically, we rely on the following cluster validation indices: Partition Coefficient (PC), Partition Entropy (PE), Modified Partition Coefficient, (MPC), Average Silhouette Width (ASW), Average Silhouette Width Fuzzy (ASWF), and Xie-Beni (XB). 

The Partition Coefficient \citep{Bezdek81} is given by:
\[
 PC=\sum_{t=1}^{T}\sum_{j=1}^{k}u_{tj}^2/T.
\]
It  can assume values between 1/$k$ and 1, with its maximum value, as function of $k$, giving us the optimal number of clusters. 

The Partition Entropy \citep{Bezdek81}:
\[ 
PE=-\sum_{t=1}^{T}\sum_{j=1}^{k}u_{tj}\ln (u_{tj})/T, 
\]
 ranges in $[0,\ln k]$ and its lowest value provides the best number of clusters.

 The Modified Partition Coefficient \citep{dave96}
\[ 
MPC=1-\dfrac{k}{k-1}(1-PC),
\]
normalizes the PC index, so that it ranges in $[0,1]$, thus eliminating dependency on $k$. 

The Average Silhouette Width \citep{rousseeuw1987} is calculated as follows: 
\[
ASW= (1/T)\sum_{t=1}^{T}\dfrac{b_{t}-a_{t}}{max\{a_{t},b_{t}\}},
\]
 with $a_{t}$ the average distance among $y_{t}$ (belonging to the cluster $\tau$) and the other observations belonging to the same group $\tau$, while $b_{t}$ is the minimum average distance among $y_{t}$ and the observations belonging to another cluster $j\neq\tau$. It can assume values between -1 and 1, with the maximum value providing us with the best number of clusters.

The Average Silhouette Width Fuzzy \citep{campello06}, 
\[
ASWF=\dfrac{\sum_{t=1}^{T}(u_{tj_1}-u_{tj_2})^\lambda ASW_{t}}{\sum_{t=1}^{T}(u_{tj_1}-u_{tj_2})^\lambda}, 
\]
 is a weighted average of the Silhouette, where $u_{tj_1}$ and $u_{tj_2}$ are the elements of the $t$--th row of $\textbf{U}$ with first and second largest values, respectively, $ASW_{t}$ the Silhouette of the $t-th$ observation, and $\lambda\geq0$ . Notice that, as opposed to ASW, it takes into account the membership grade matrix. 

Finally, the Xie-Beni index \citep{xie91}, 
\[
XB=\dfrac{\sum_{t=1}^{T}\sum_{j=1}^{k}u^2_{tj}d^2(y_{t},c_{j})}{T \min\limits_{i,j} d^2({c_{i},c_{j}})},
\]
 is minimized to obtain the best partition. Notice that the XB index takes into account both membership grade matrix and the observations.

\section{Monte Carlo Evidence}\label{sec:mc}
The DGPs used for the Monte Carlo experiments are the models in equation  (\ref{ms_2})  (call it MS(2)), equation (\ref{ms_3}) (MS(3)) and MS--AR(1) models like (\ref{ms_ar}), with $p=1$, and with 2 (MS(2)-AR(1)) or 3 (MS(3)-AR(1)) states, adopting the Normal distribution for $\varepsilon_t$. We cover several scenarios, combining a set of parameters that yields the 32 models (8 of each model type) shown in Table \ref{tab:mods} (labeled to identify them).

\begin{figure}[H]
	\caption{Mixture density functions relative to 32 DGPs described in Table \ref{tab:mods}, with weights equal to the corresponding ergodic probabilities.}
	\begin{subfigure}[h!]{0.24\textwidth}
		\centering		  
		\includegraphics[width=\textwidth]{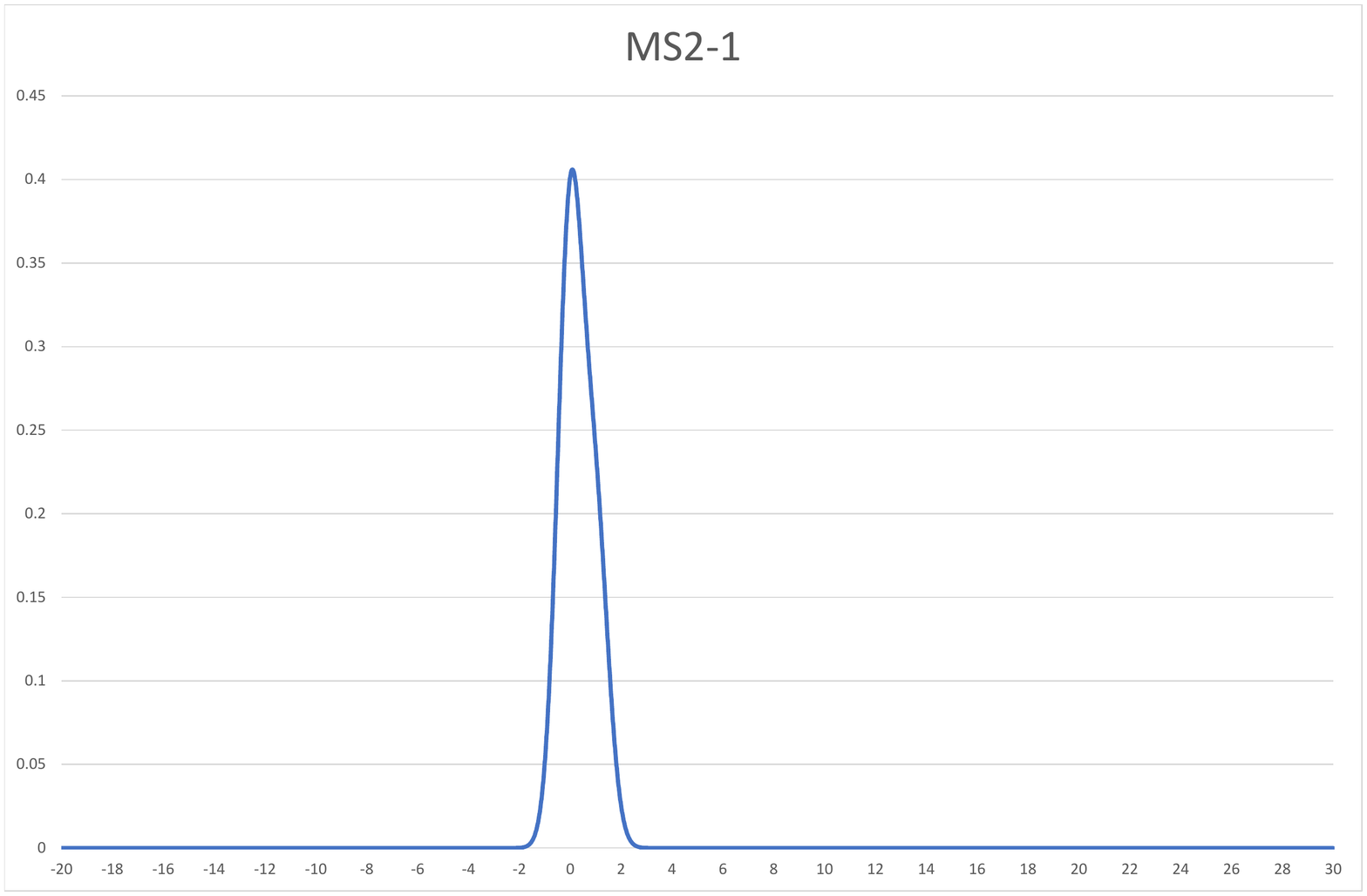}  
	\end{subfigure}
	\hfill 
	\begin{subfigure}[h!]{0.24\textwidth}
		\centering
		\includegraphics[width=\textwidth]{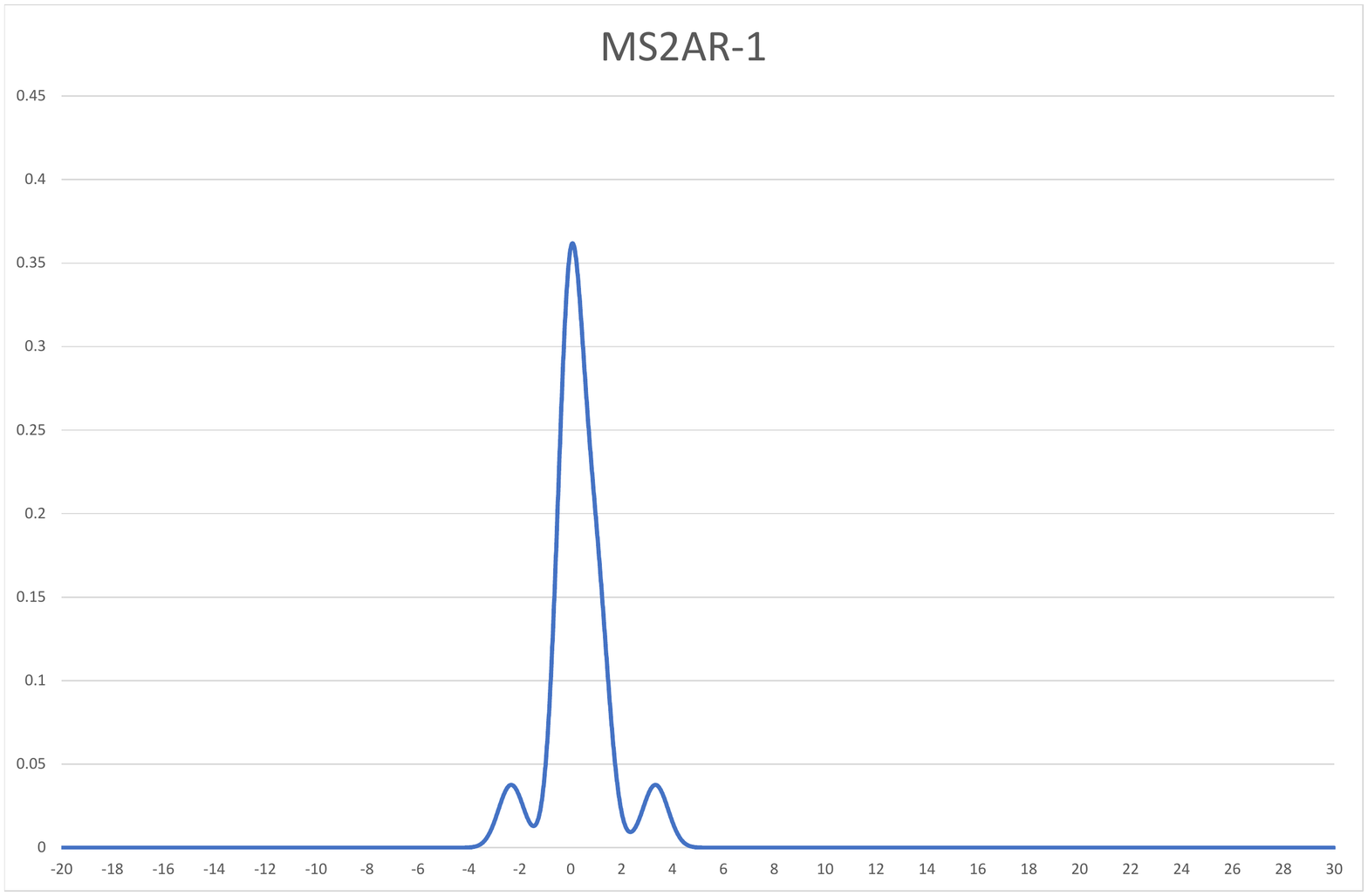}
	\end{subfigure}   
	\hfill
	\begin{subfigure}[h!]{0.24\textwidth}
		\centering		
		\includegraphics[width=\textwidth]{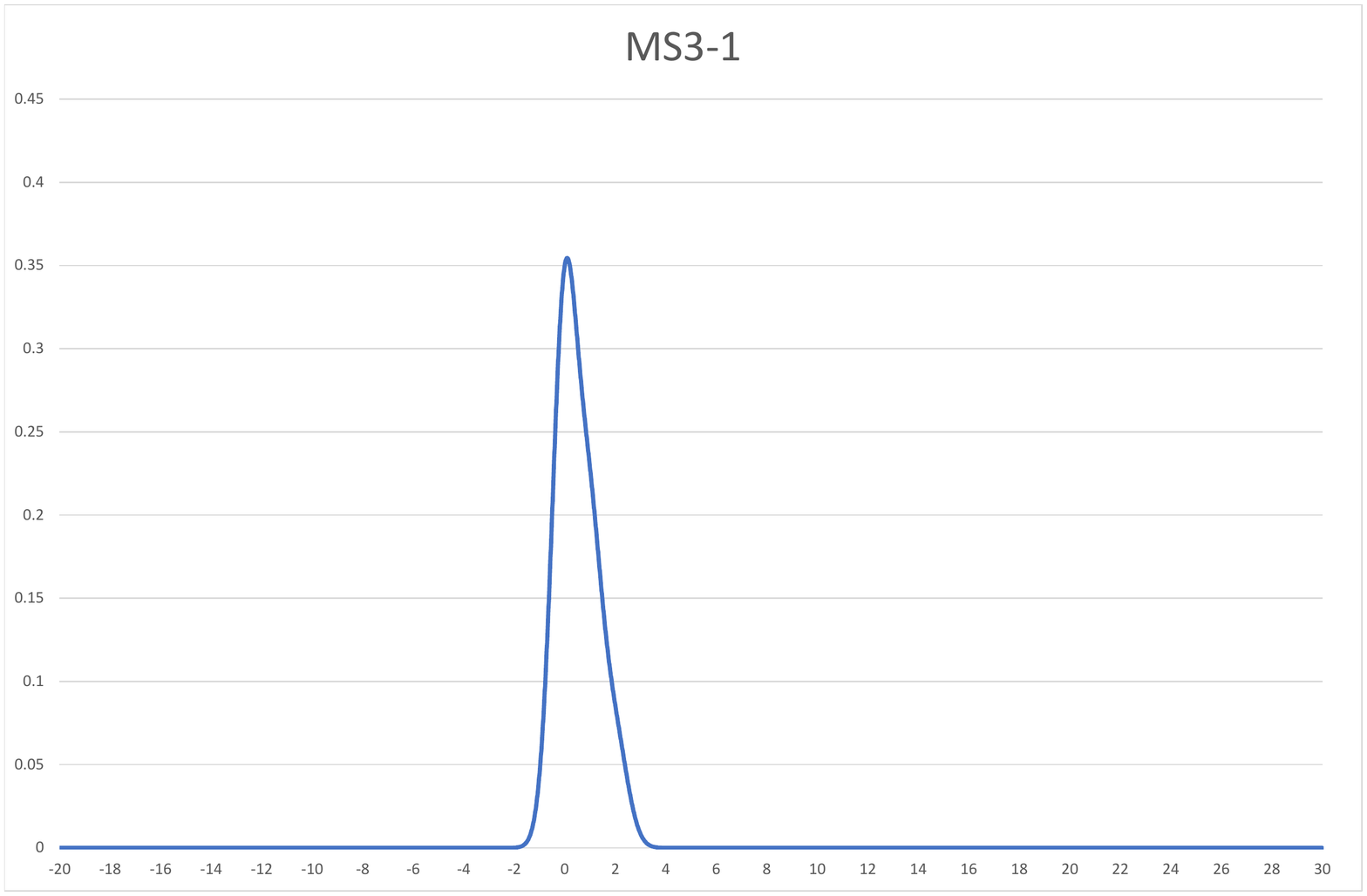} 
	\end{subfigure} 
	\hfill
	\begin{subfigure}[h!]{0.24\textwidth}
		\centering		
		\includegraphics[width=\textwidth]{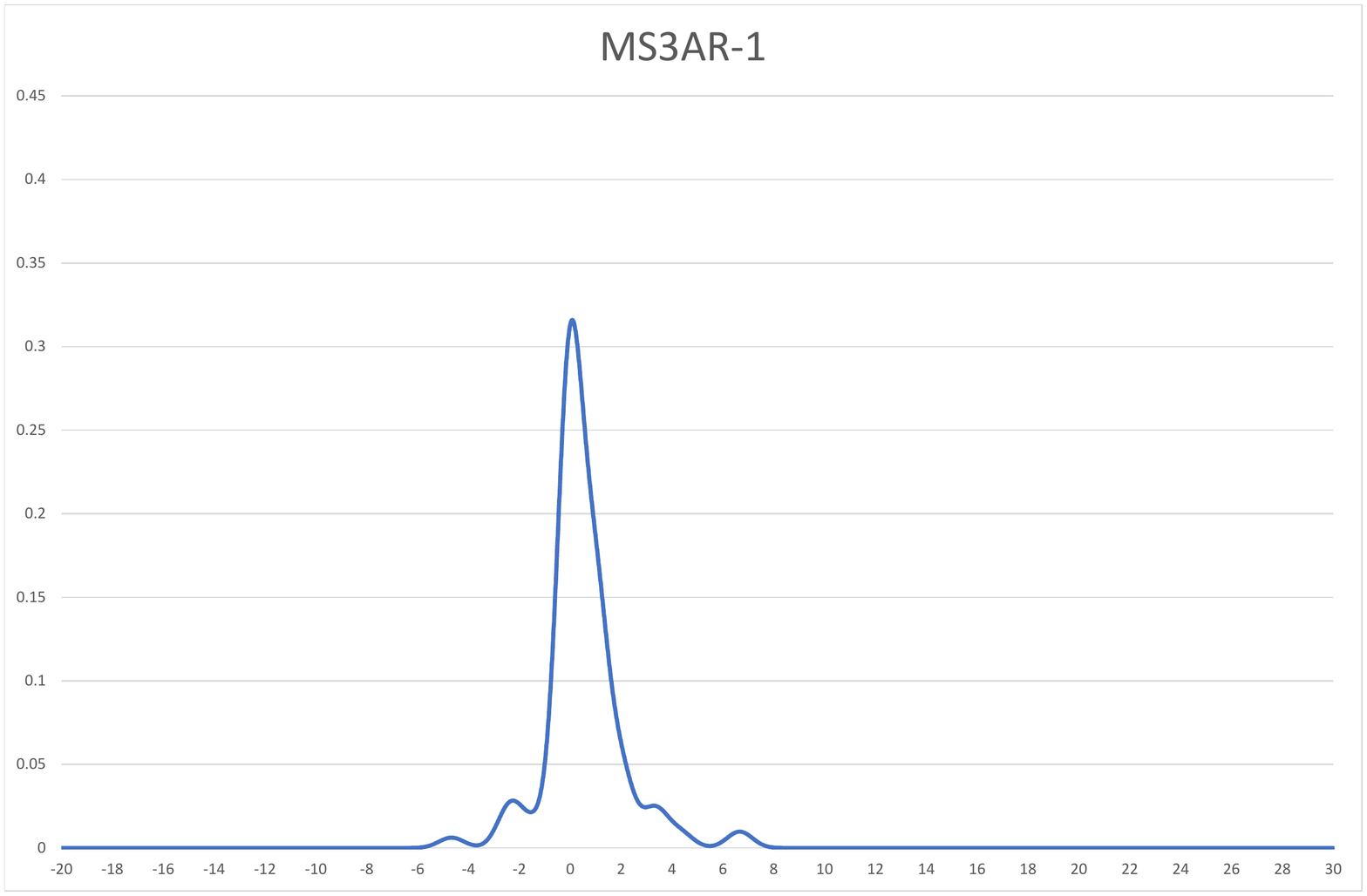} 
	\end{subfigure} 
	\vfill
	\begin{subfigure}[h!]{0.24\textwidth}
		\centering		  
		\includegraphics[width=\textwidth]{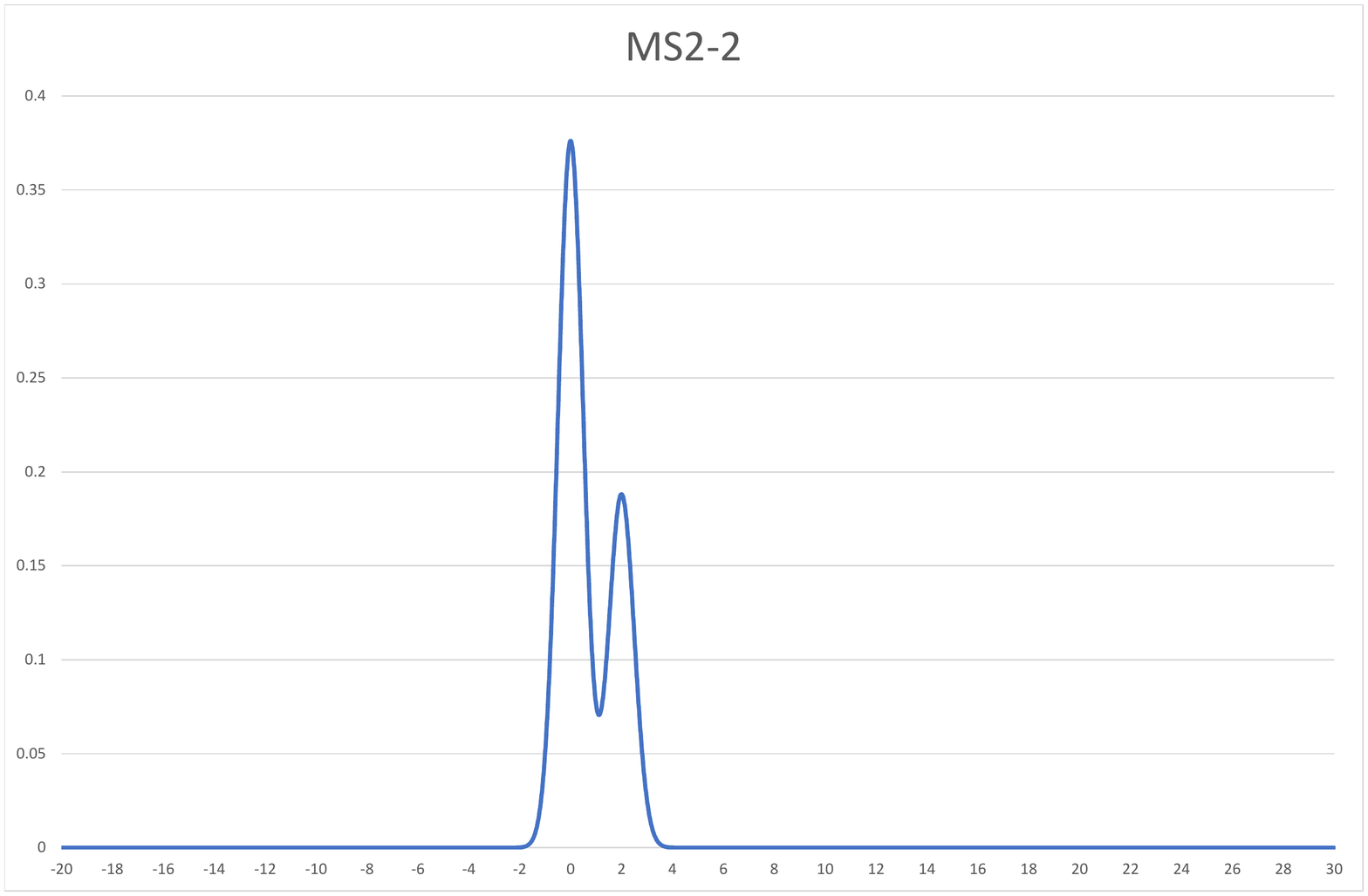}  
	\end{subfigure}
	\hfill 
	\begin{subfigure}[h!]{0.24\textwidth}
		\centering
		\includegraphics[width=\textwidth]{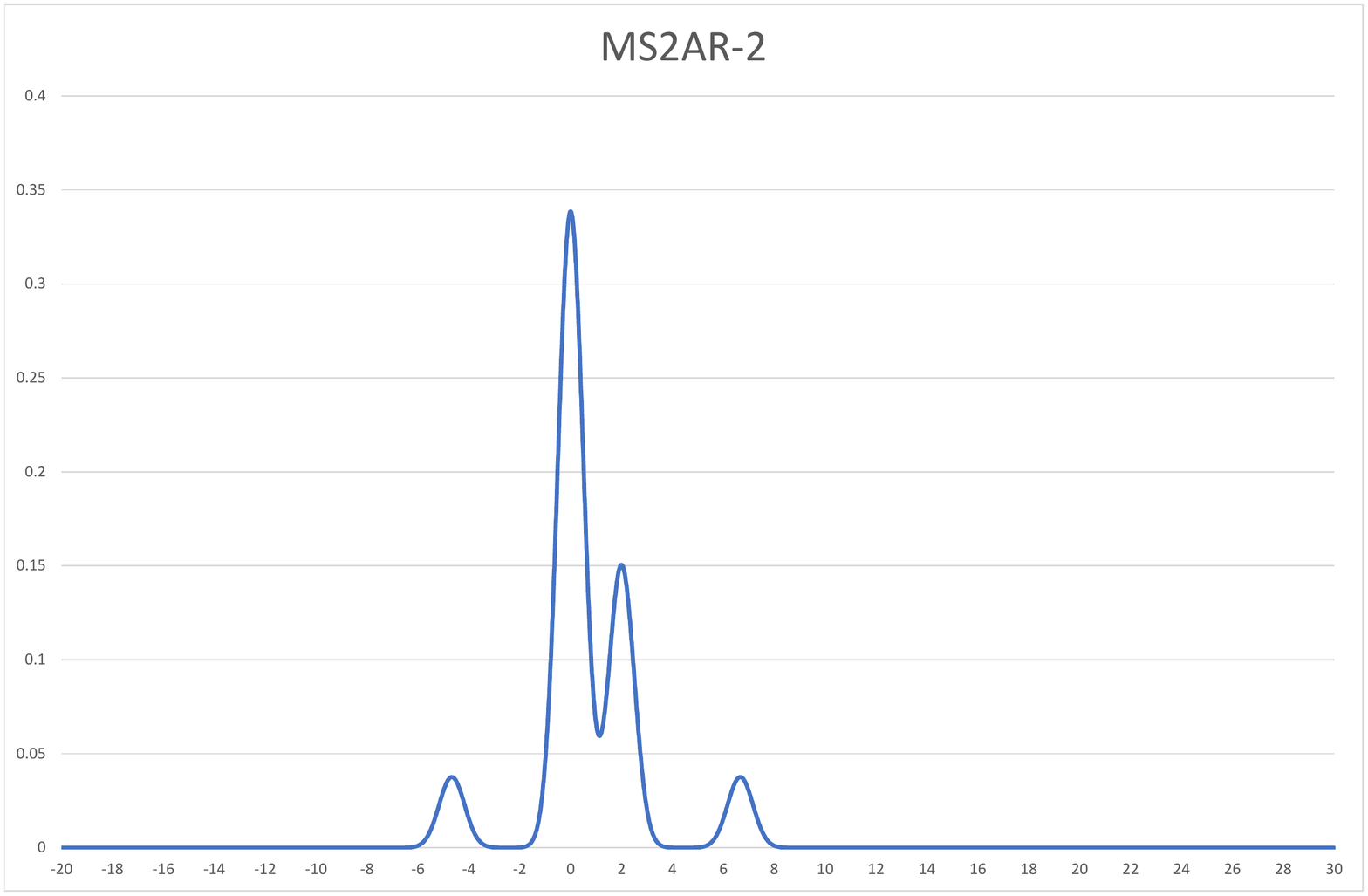}
	\end{subfigure}   
	\hfill
	\begin{subfigure}[h!]{0.24\textwidth}
		\centering		
		\includegraphics[width=\textwidth]{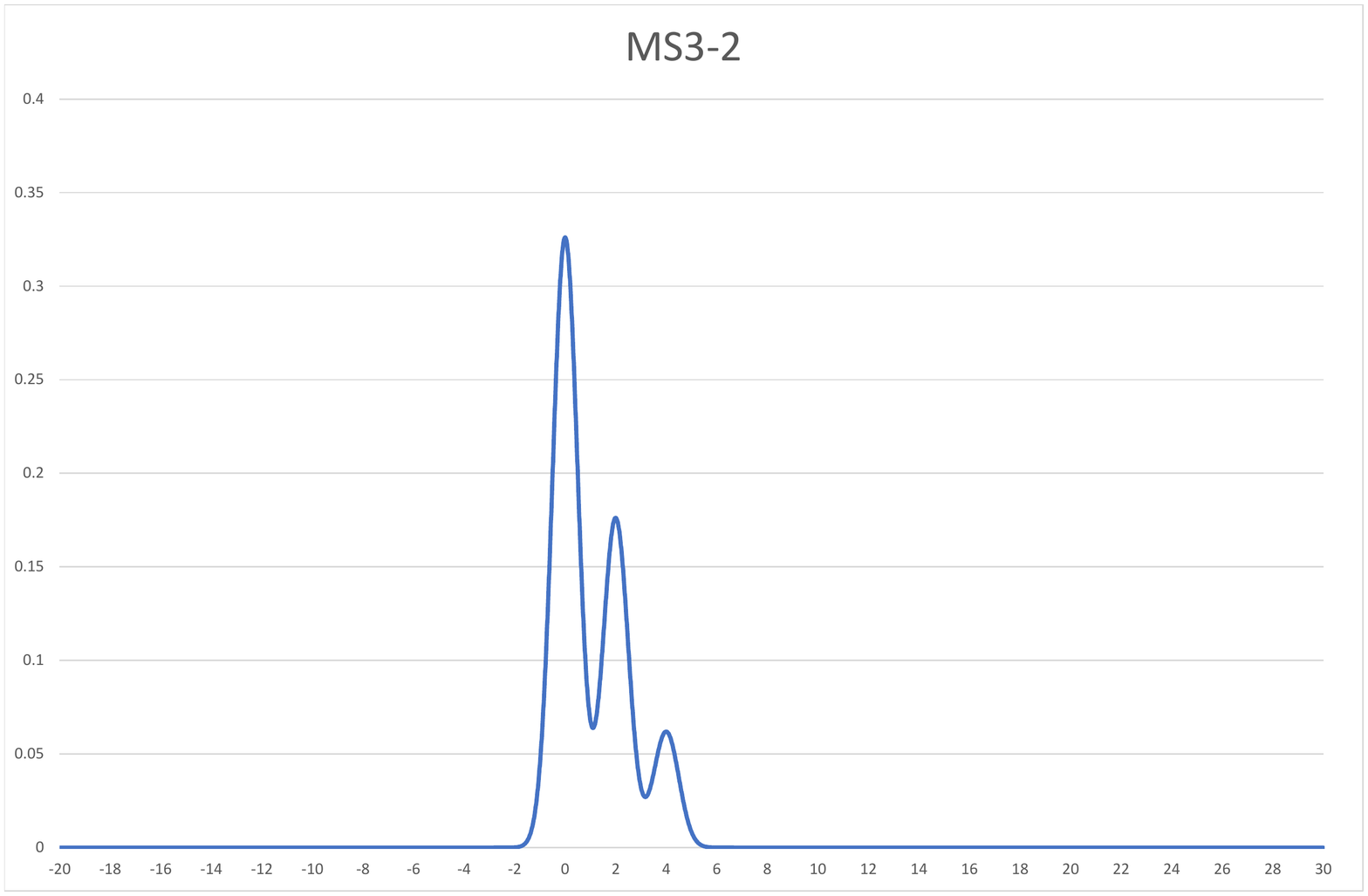} 
	\end{subfigure} 
	\hfill
	\begin{subfigure}[h!]{0.24\textwidth}
		\centering		
		\includegraphics[width=\textwidth]{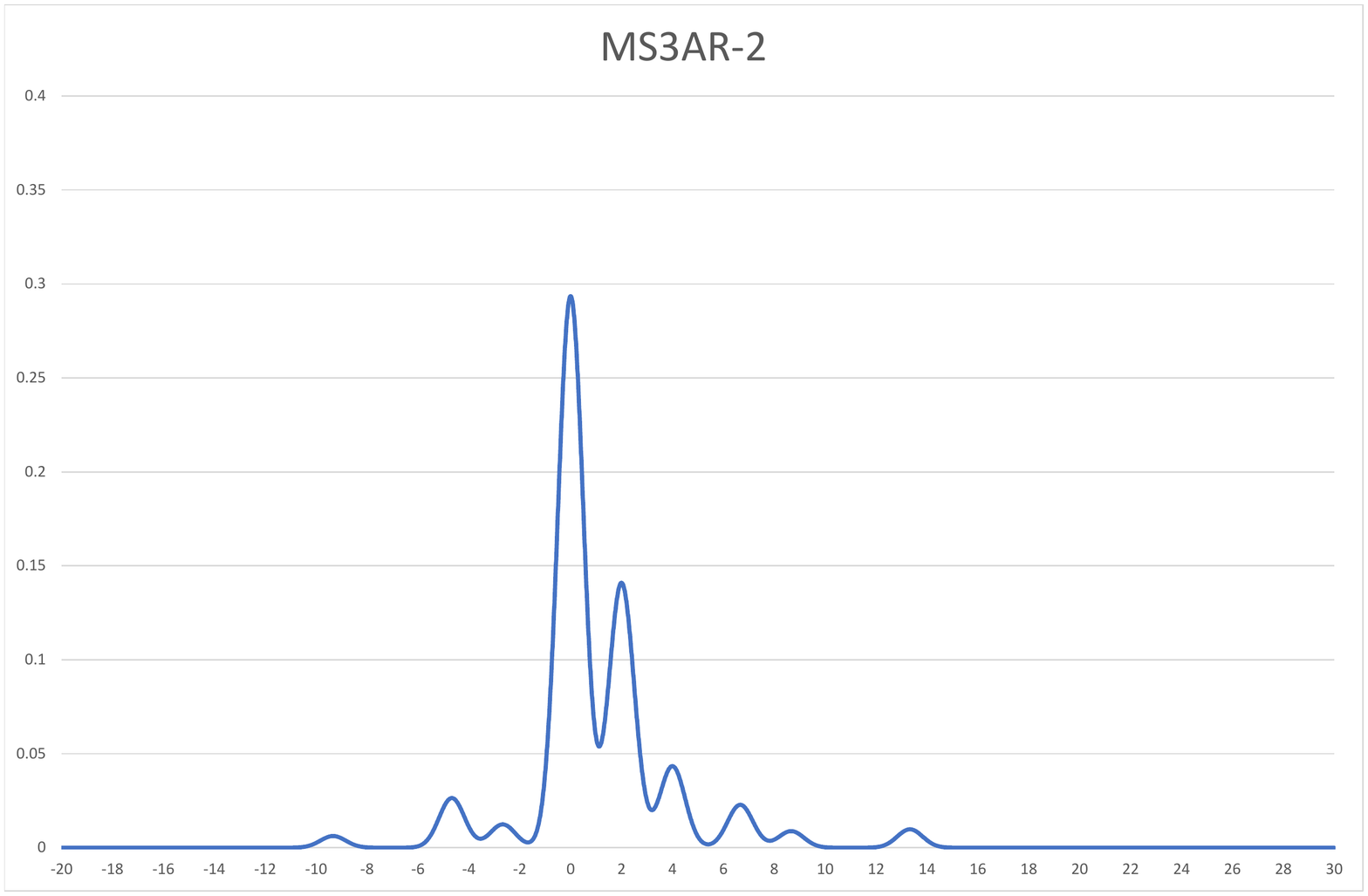} 
	\end{subfigure} 
	\vfill
	\begin{subfigure}[h!]{0.24\textwidth}
		\centering		  
		\includegraphics[width=\textwidth]{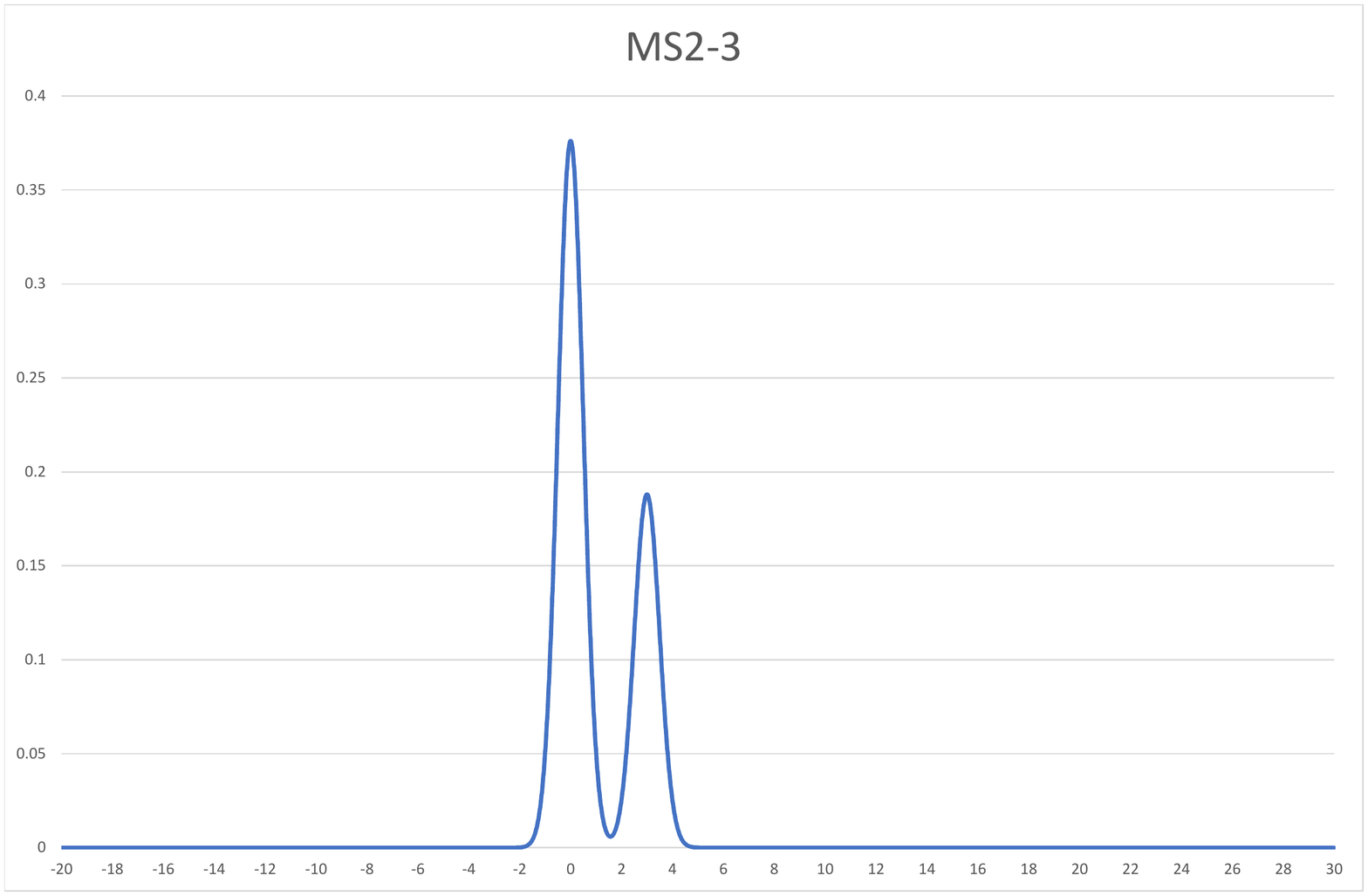}  
	\end{subfigure}
	\hfill 
	\begin{subfigure}[h!]{0.24\textwidth}
		\centering
		\includegraphics[width=\textwidth]{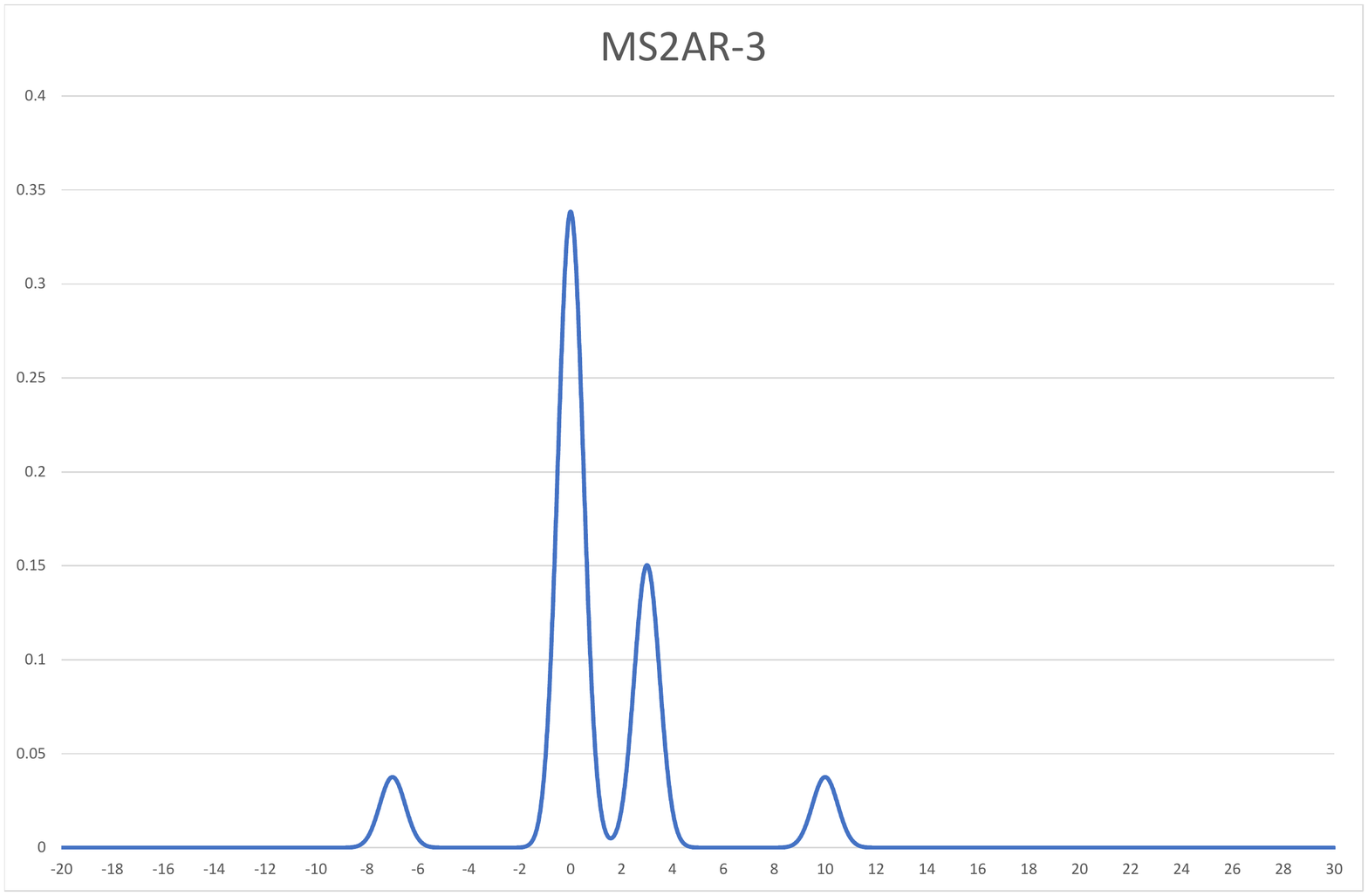}
	\end{subfigure}   
	\hfill
	\begin{subfigure}[h!]{0.24\textwidth}
		\centering		
		\includegraphics[width=\textwidth]{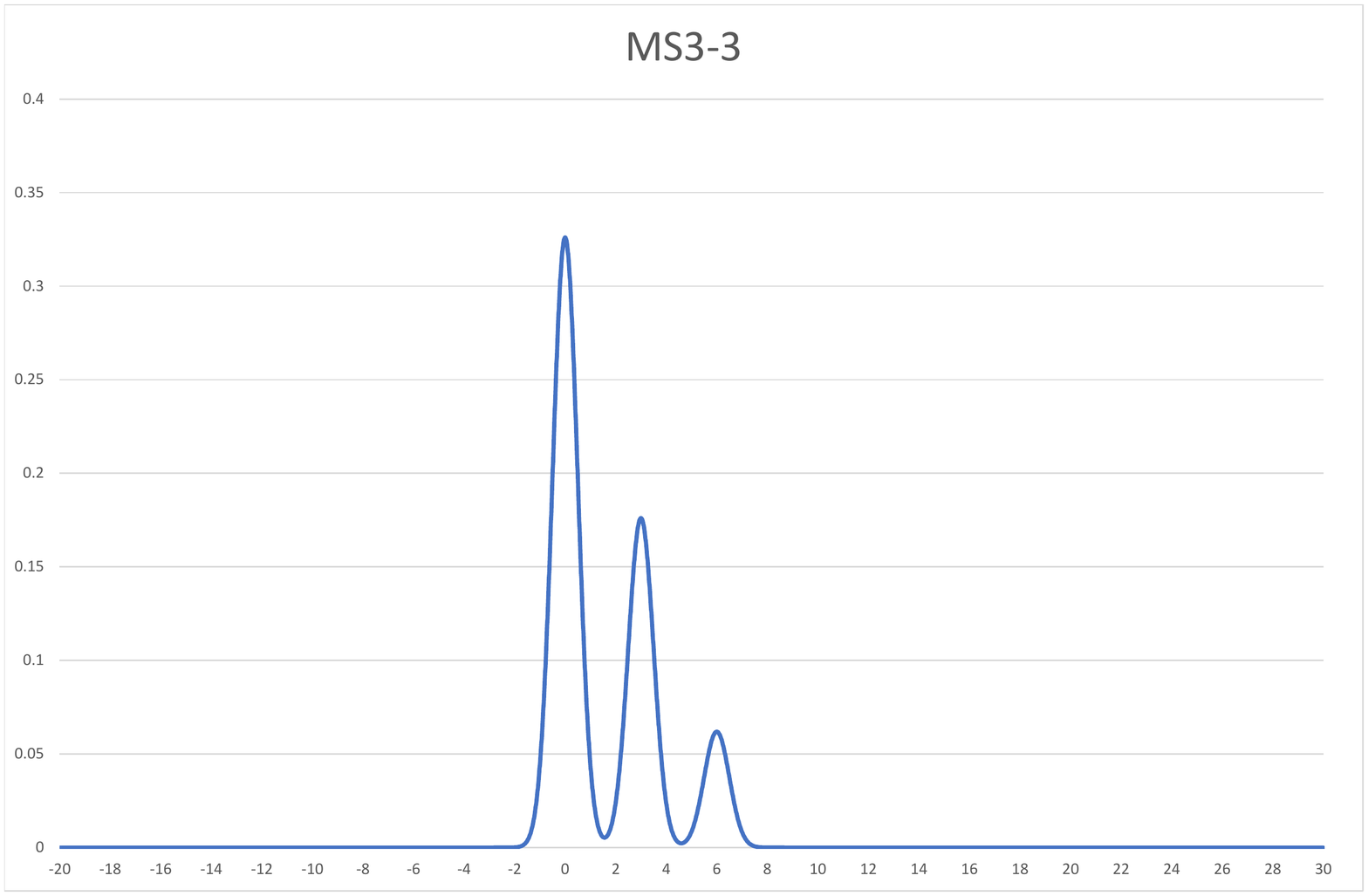} 
	\end{subfigure} 
	\hfill
	\begin{subfigure}[h!]{0.24\textwidth}
		\centering		
		\includegraphics[width=\textwidth]{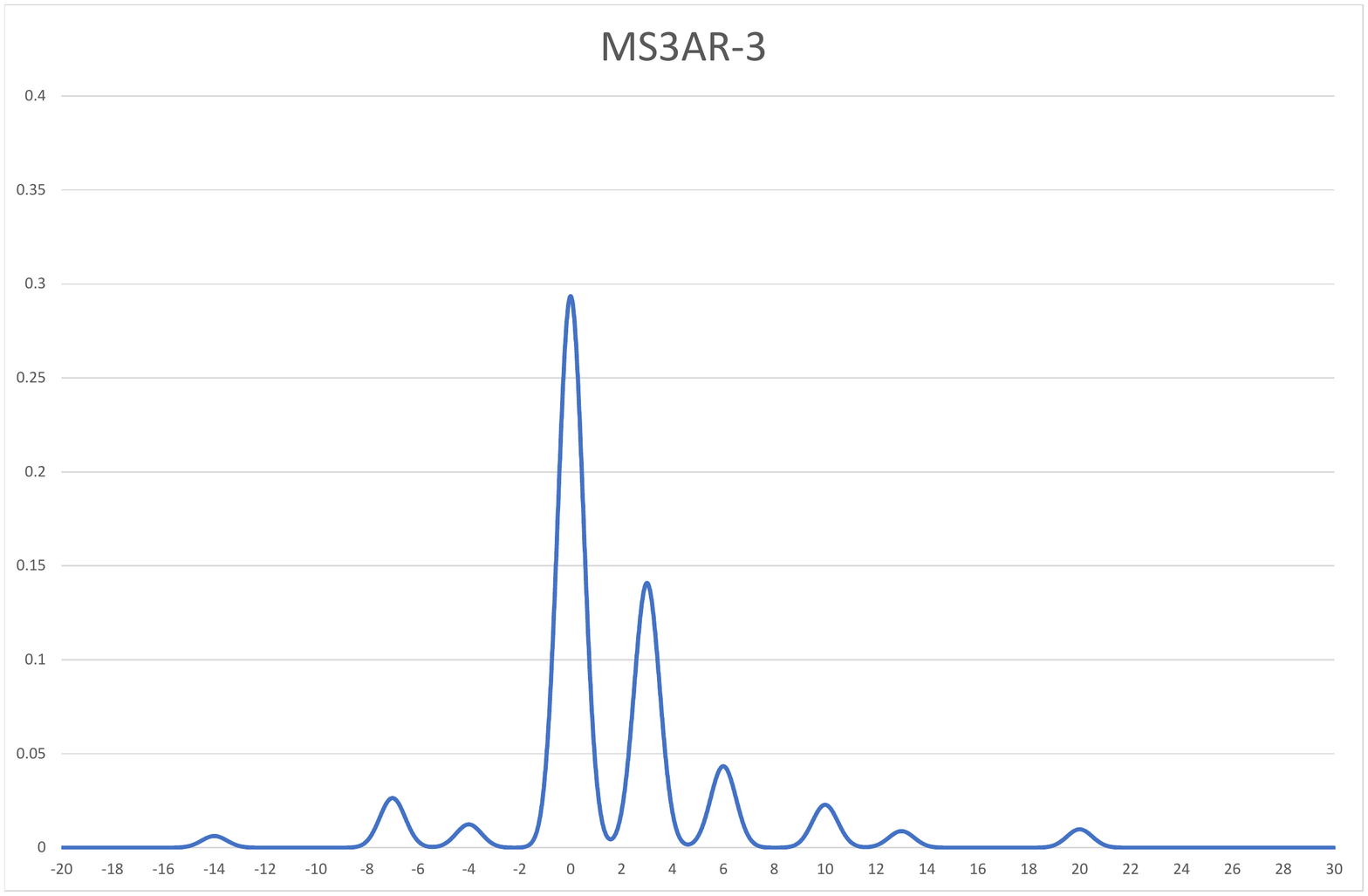} 
	\end{subfigure} 
	\vfill
	\begin{subfigure}[h!]{0.24\textwidth}
		\centering		  
		\includegraphics[width=\textwidth]{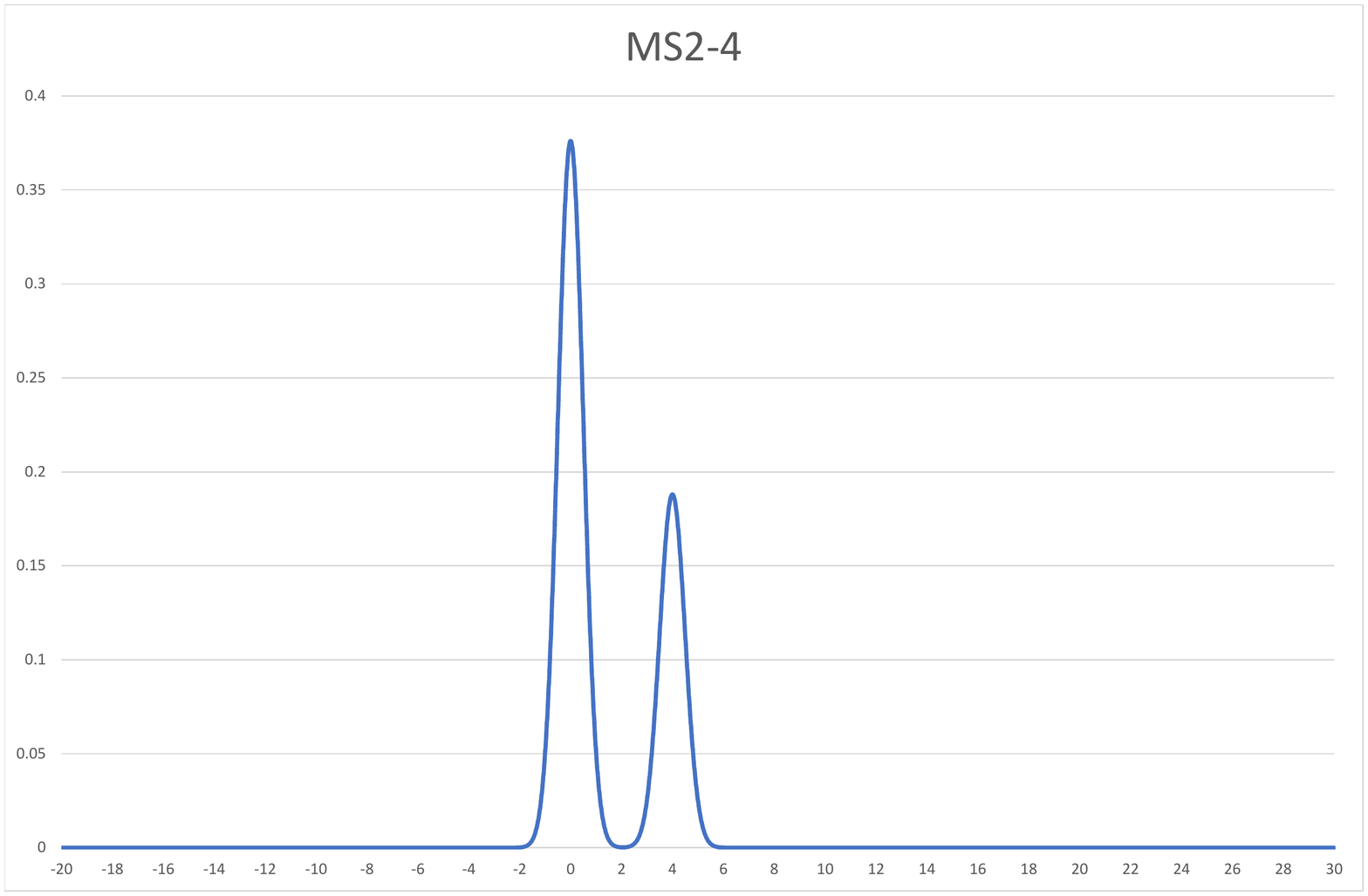}  
	\end{subfigure}
	\hfill 
	\begin{subfigure}[h!]{0.24\textwidth}
		\centering
		\includegraphics[width=\textwidth]{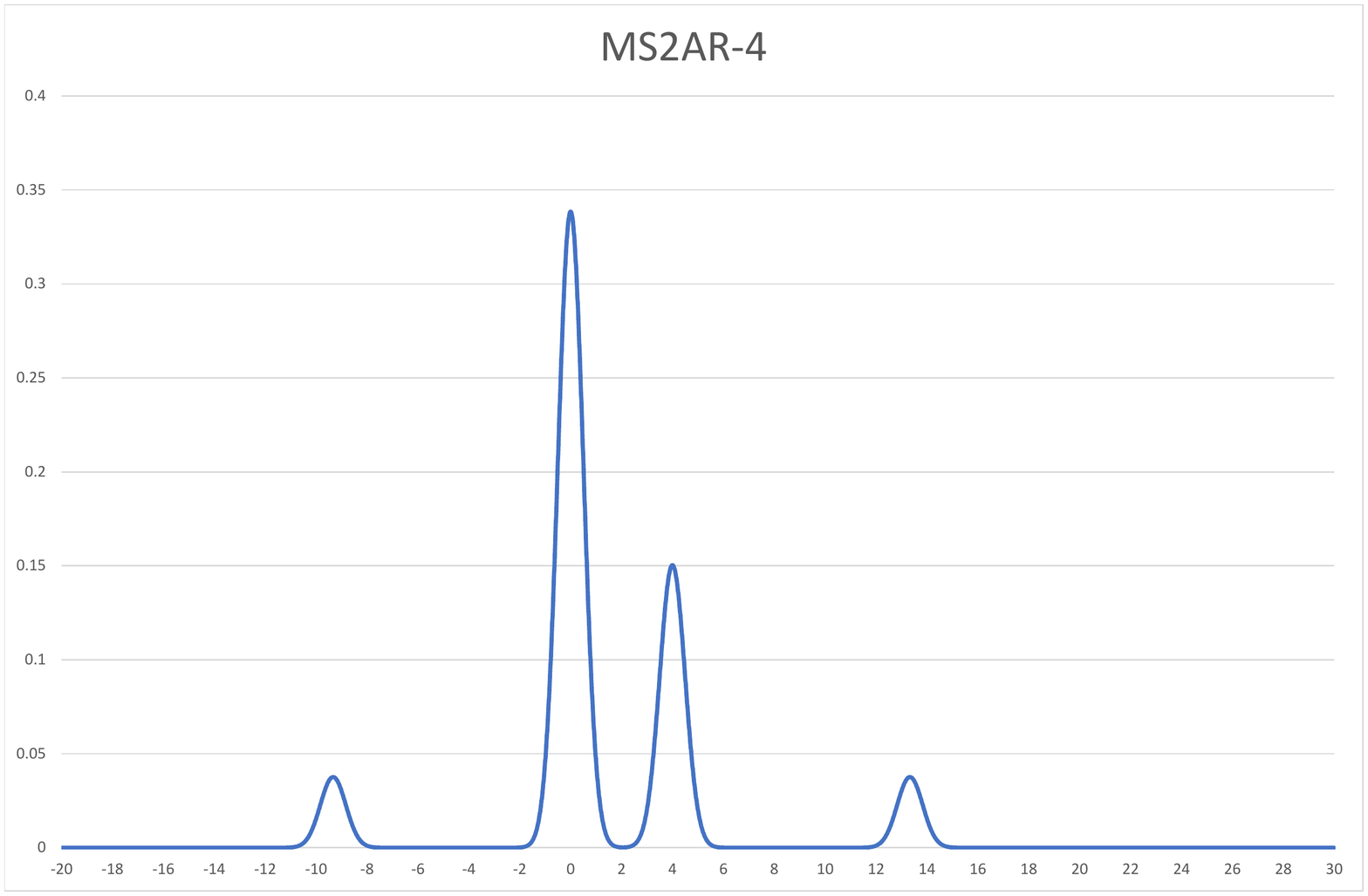}
	\end{subfigure}   
	\hfill
	\begin{subfigure}[h!]{0.24\textwidth}
		\centering		
		\includegraphics[width=\textwidth]{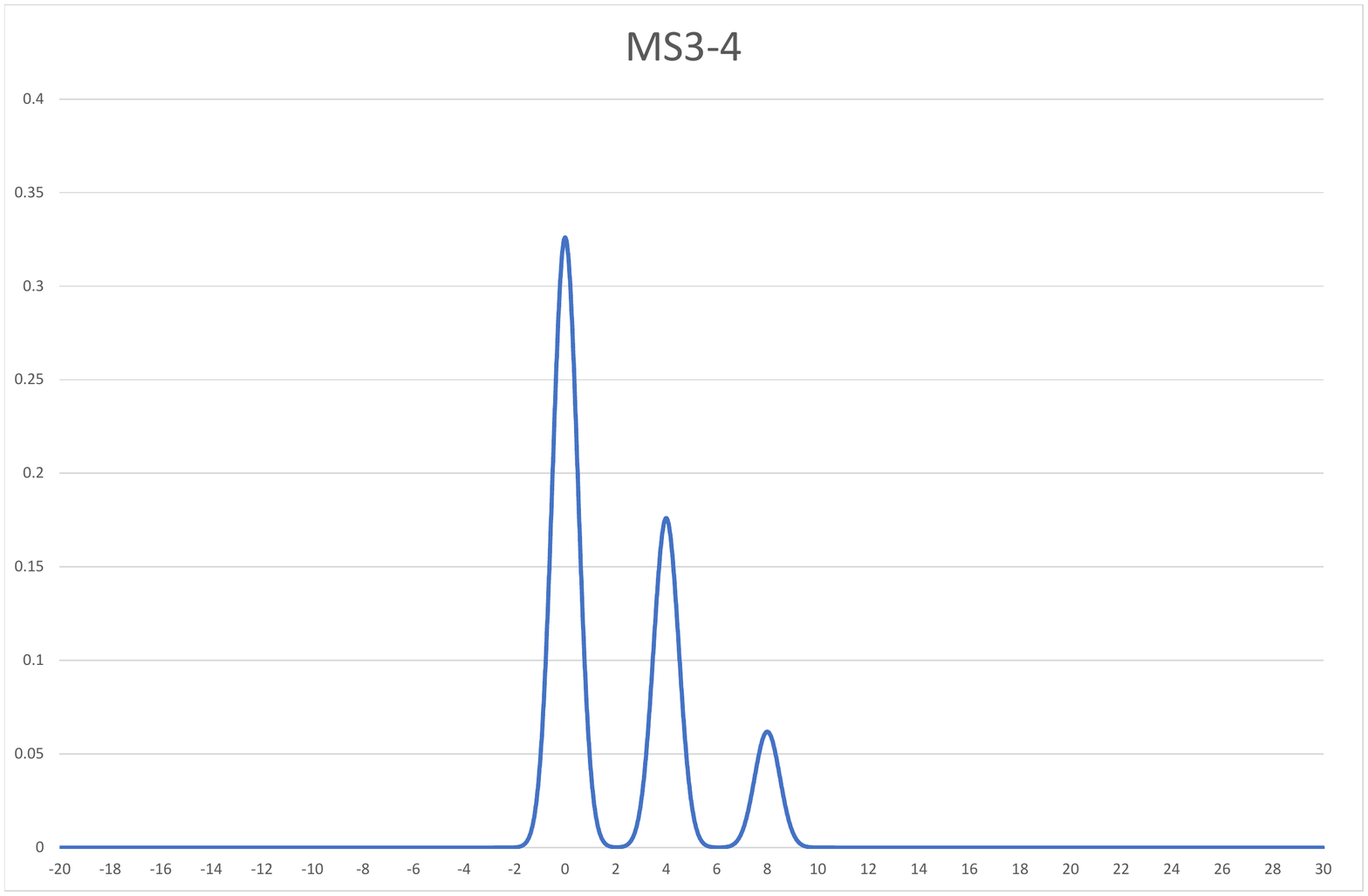} 
	\end{subfigure} 
	\hfill
	\begin{subfigure}[h!]{0.24\textwidth}
		\centering		
		\includegraphics[width=\textwidth]{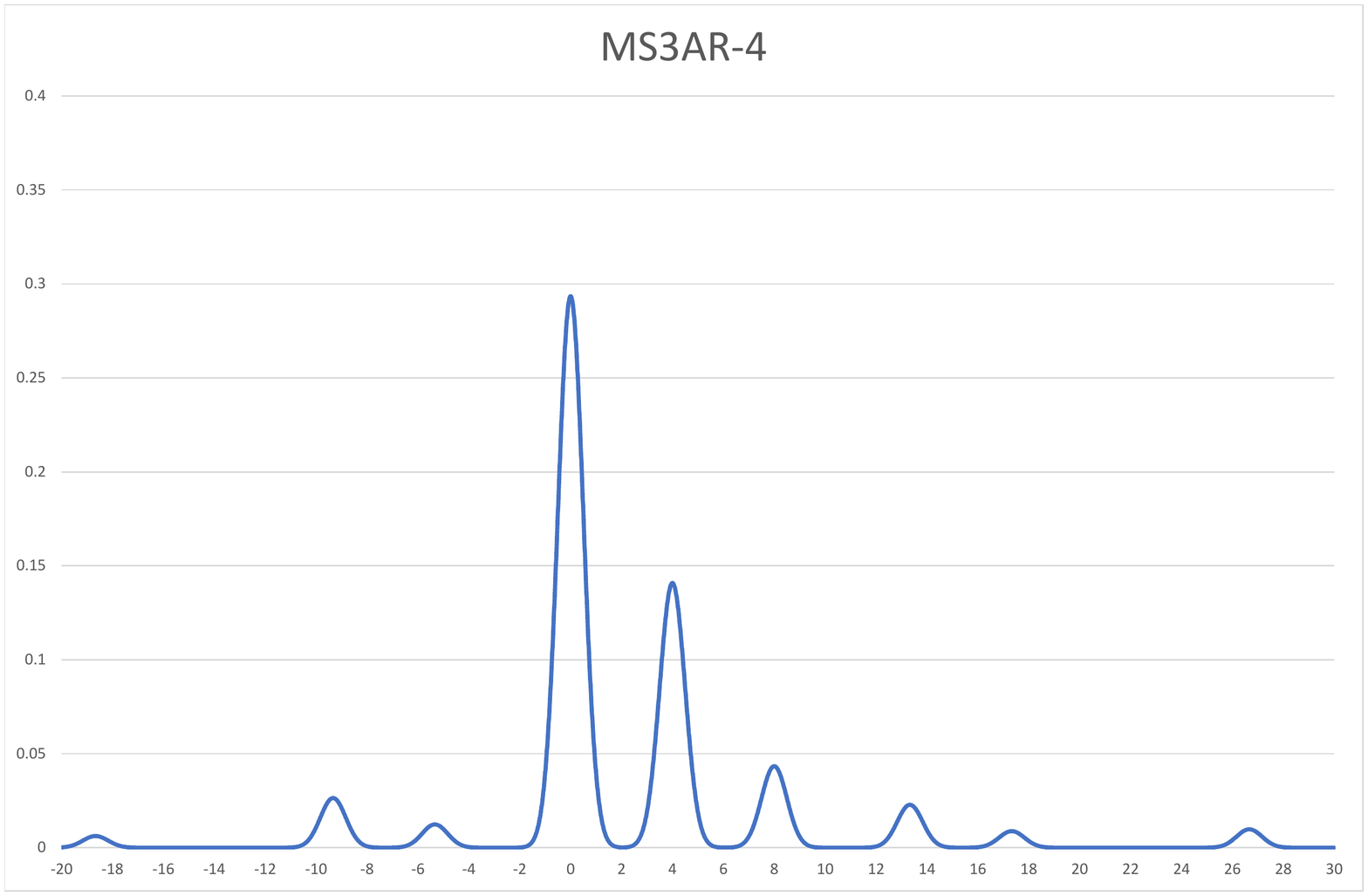} 
	\end{subfigure} 
	\vfill
	\begin{subfigure}[h!]{0.24\textwidth}
		\centering		  
		\includegraphics[width=\textwidth]{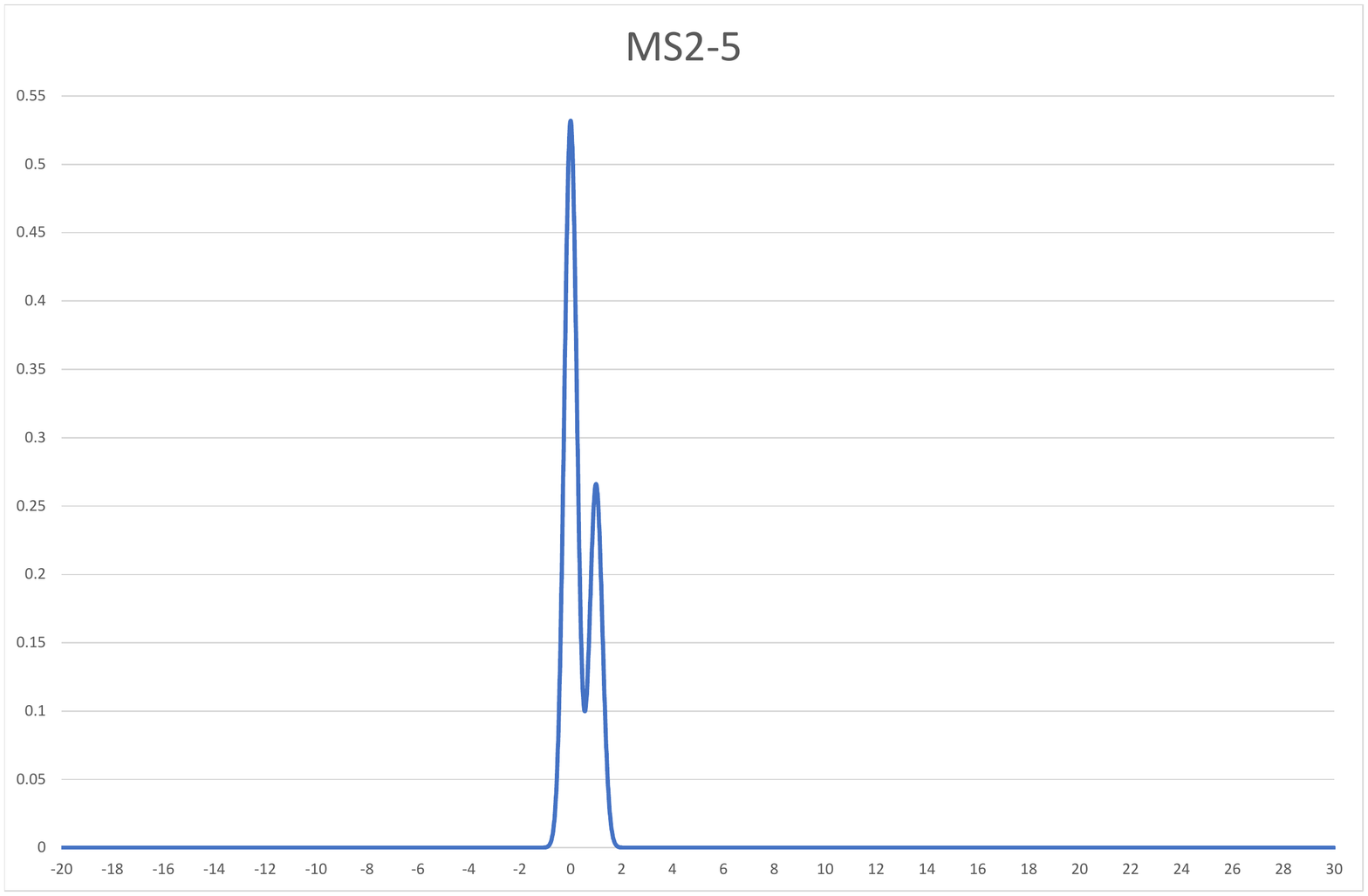}  
	\end{subfigure}
	\hfill 
	\begin{subfigure}[h!]{0.24\textwidth}
		\centering
		\includegraphics[width=\textwidth]{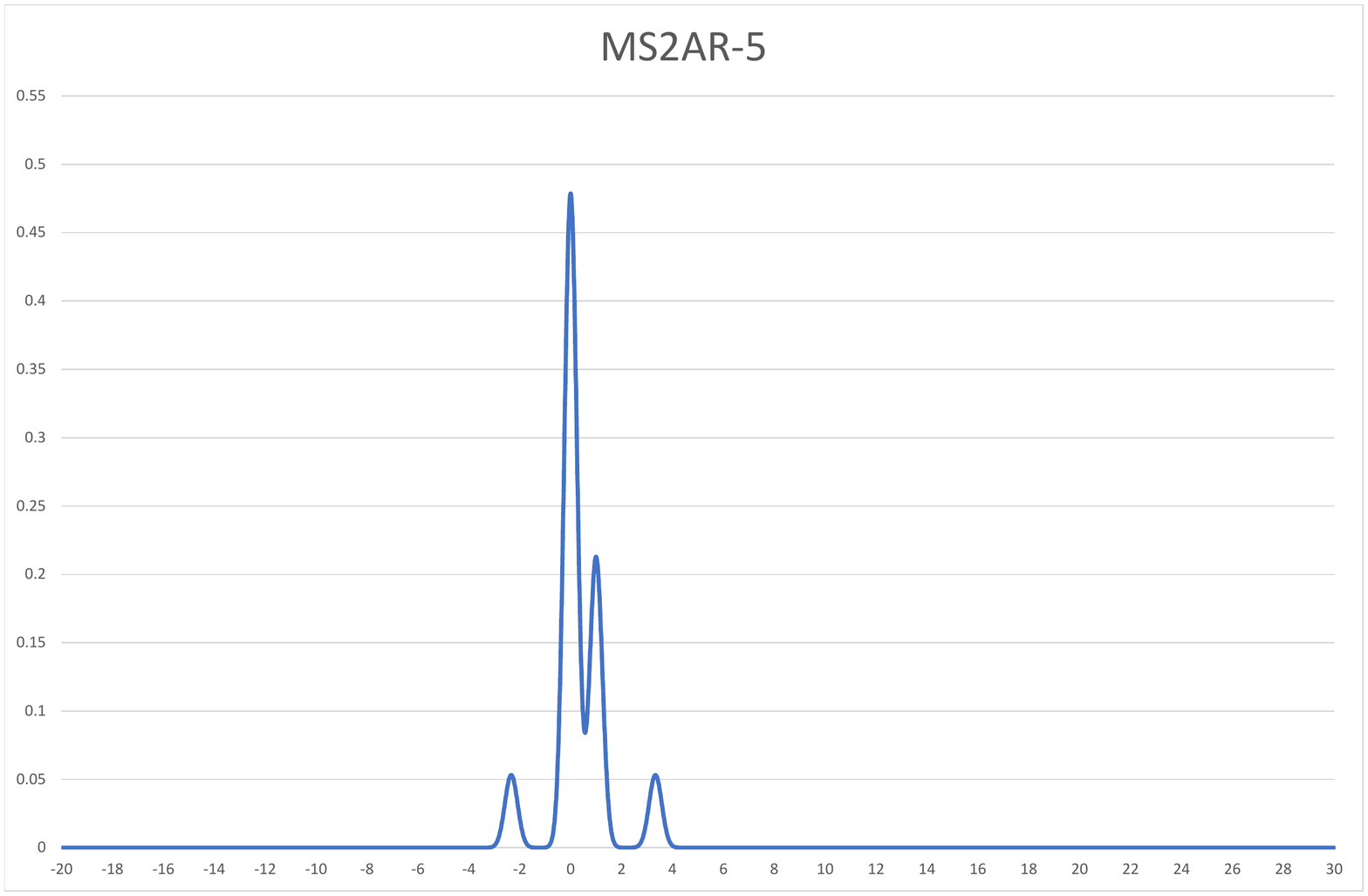}
	\end{subfigure}   
	\hfill
	\begin{subfigure}[h!]{0.24\textwidth}
		\centering		
		\includegraphics[width=\textwidth]{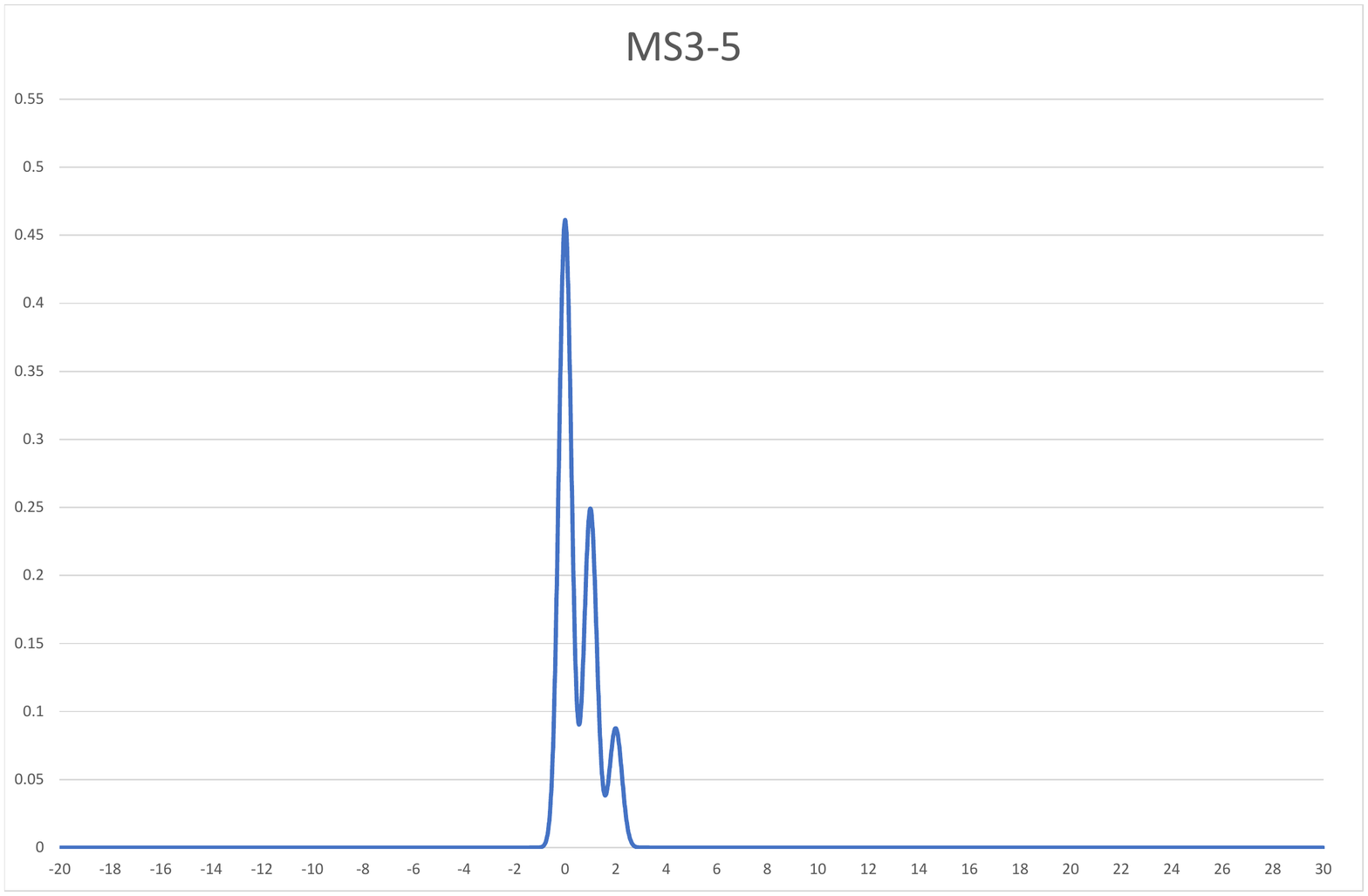} 
	\end{subfigure} 
	\hfill
	\begin{subfigure}[h!]{0.24\textwidth}
		\centering		
		\includegraphics[width=\textwidth]{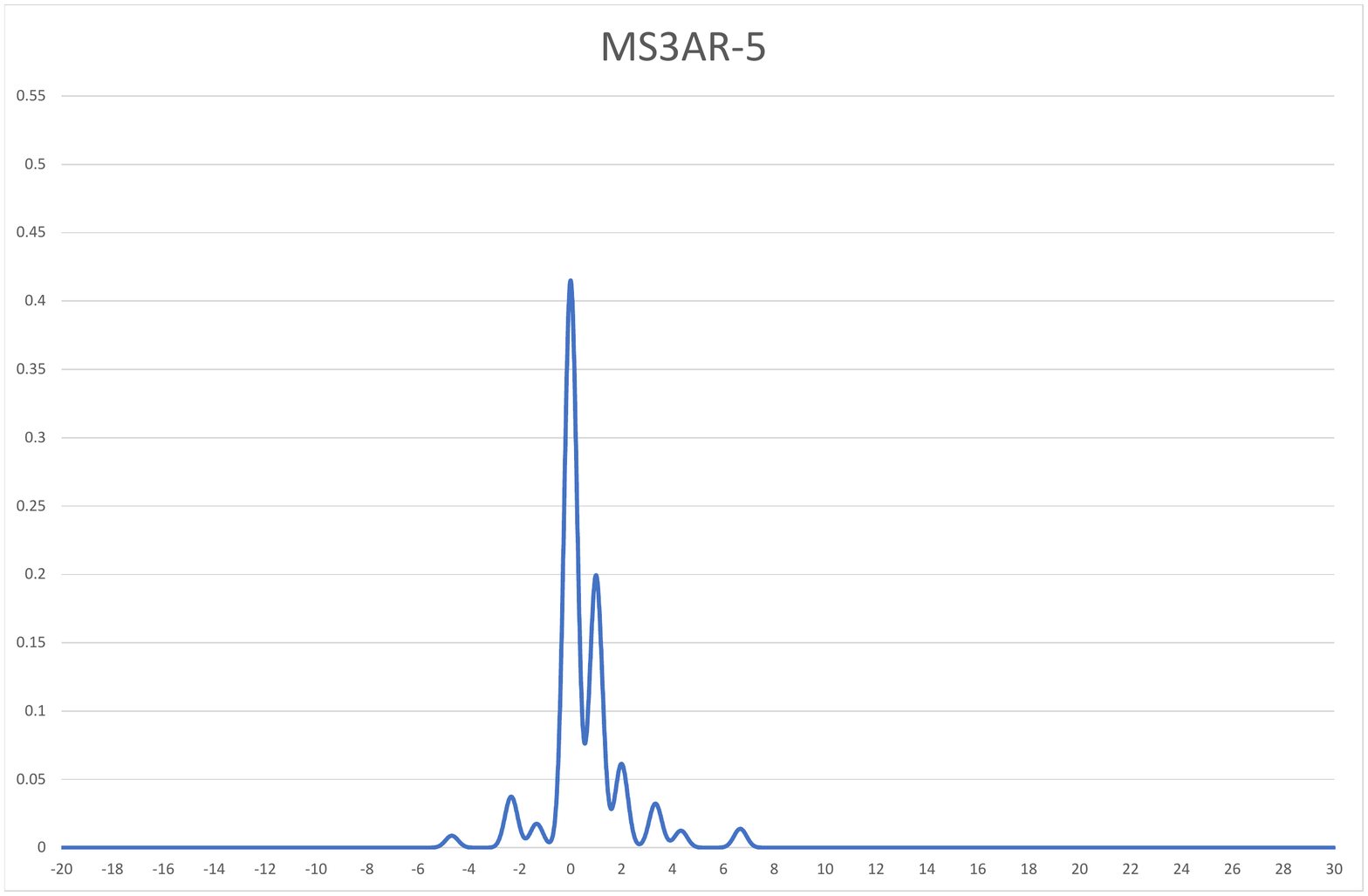} 
	\end{subfigure} 
	\vfill
	\begin{subfigure}[h!]{0.24\textwidth}
		\centering		  
		\includegraphics[width=\textwidth]{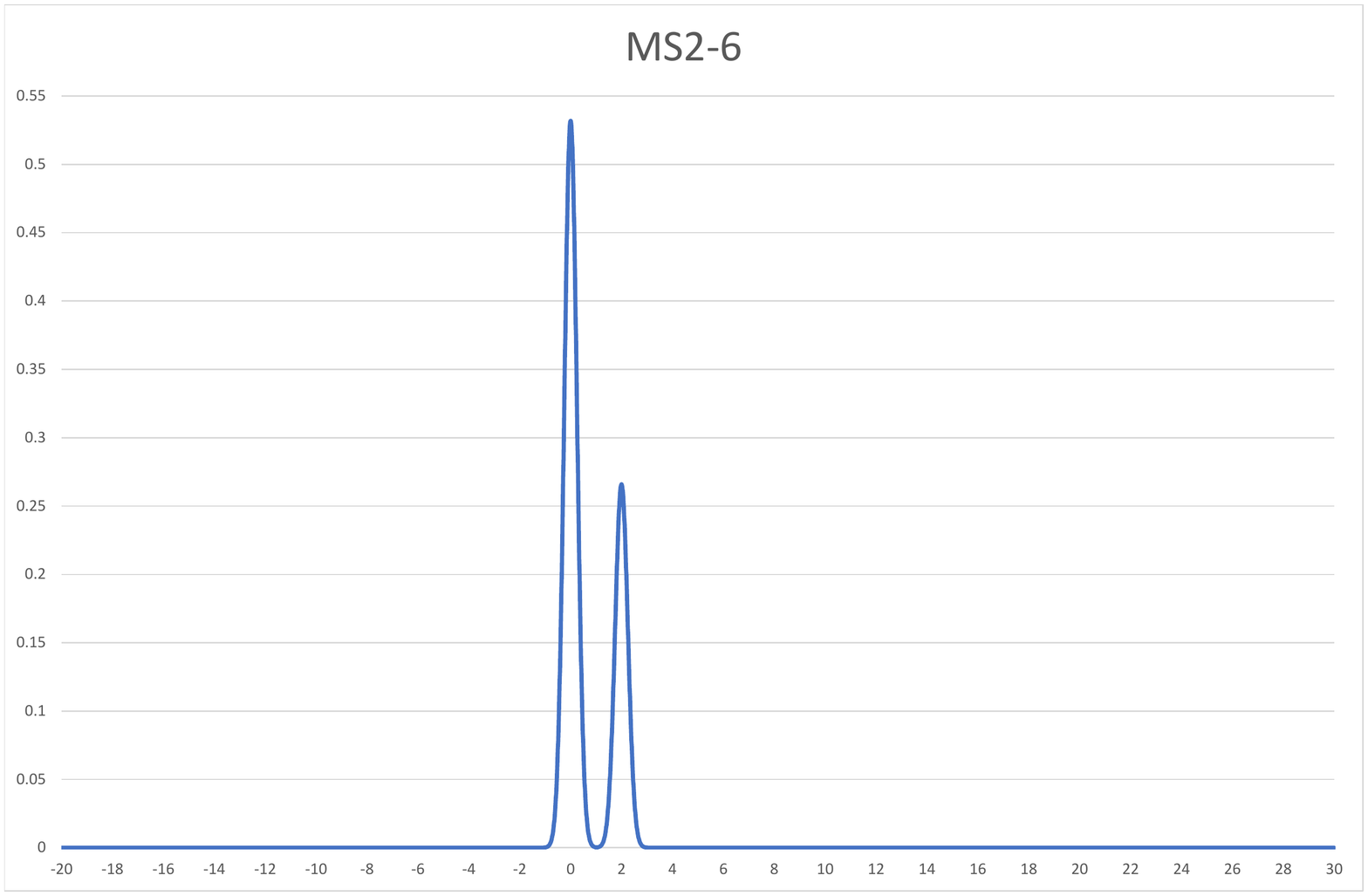}  
	\end{subfigure}
	\hfill 
	\begin{subfigure}[h!]{0.24\textwidth}
		\centering
		\includegraphics[width=\textwidth]{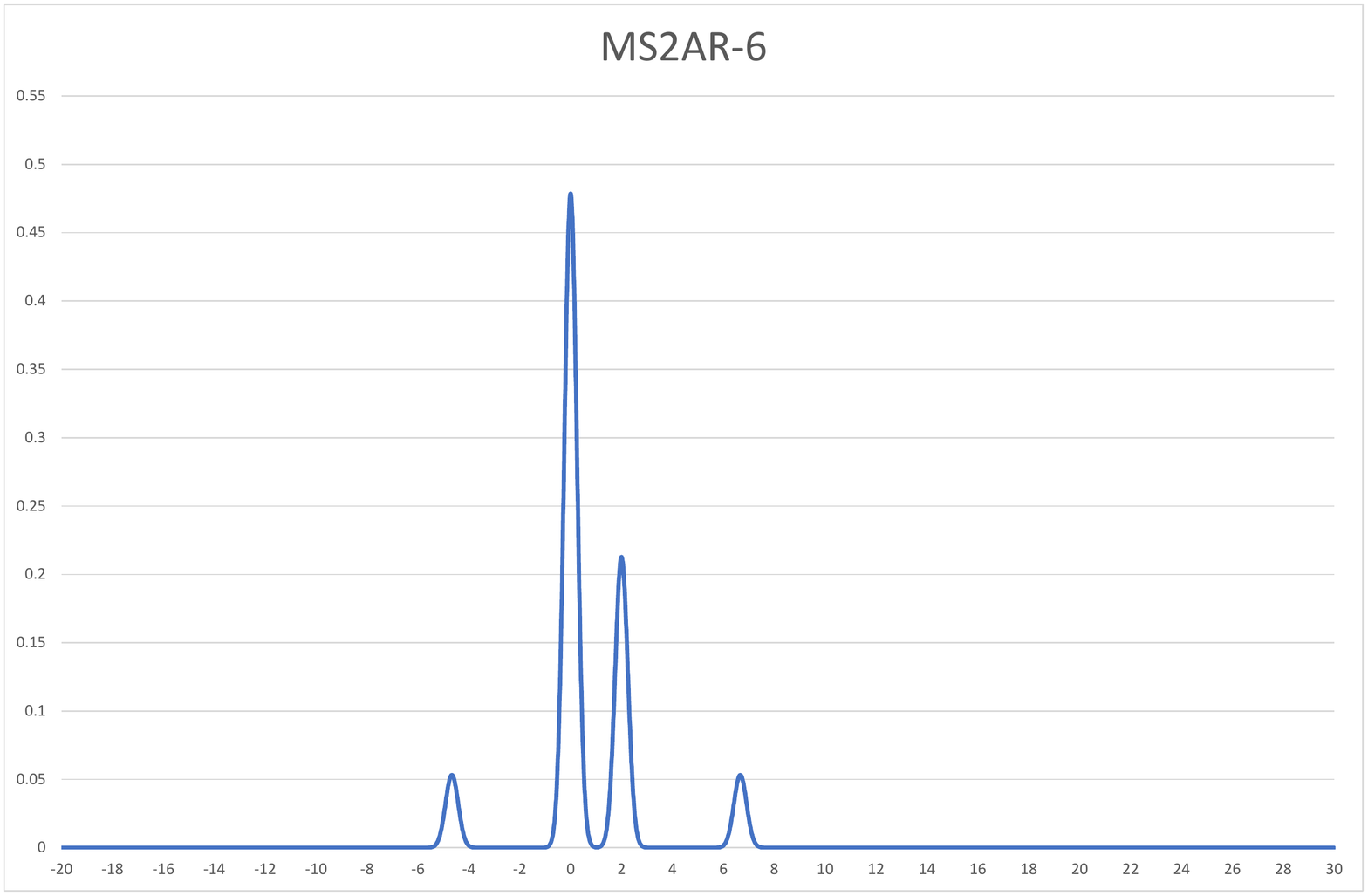}
	\end{subfigure}   
	\hfill
	\begin{subfigure}[h!]{0.24\textwidth}
		\centering		
		\includegraphics[width=\textwidth]{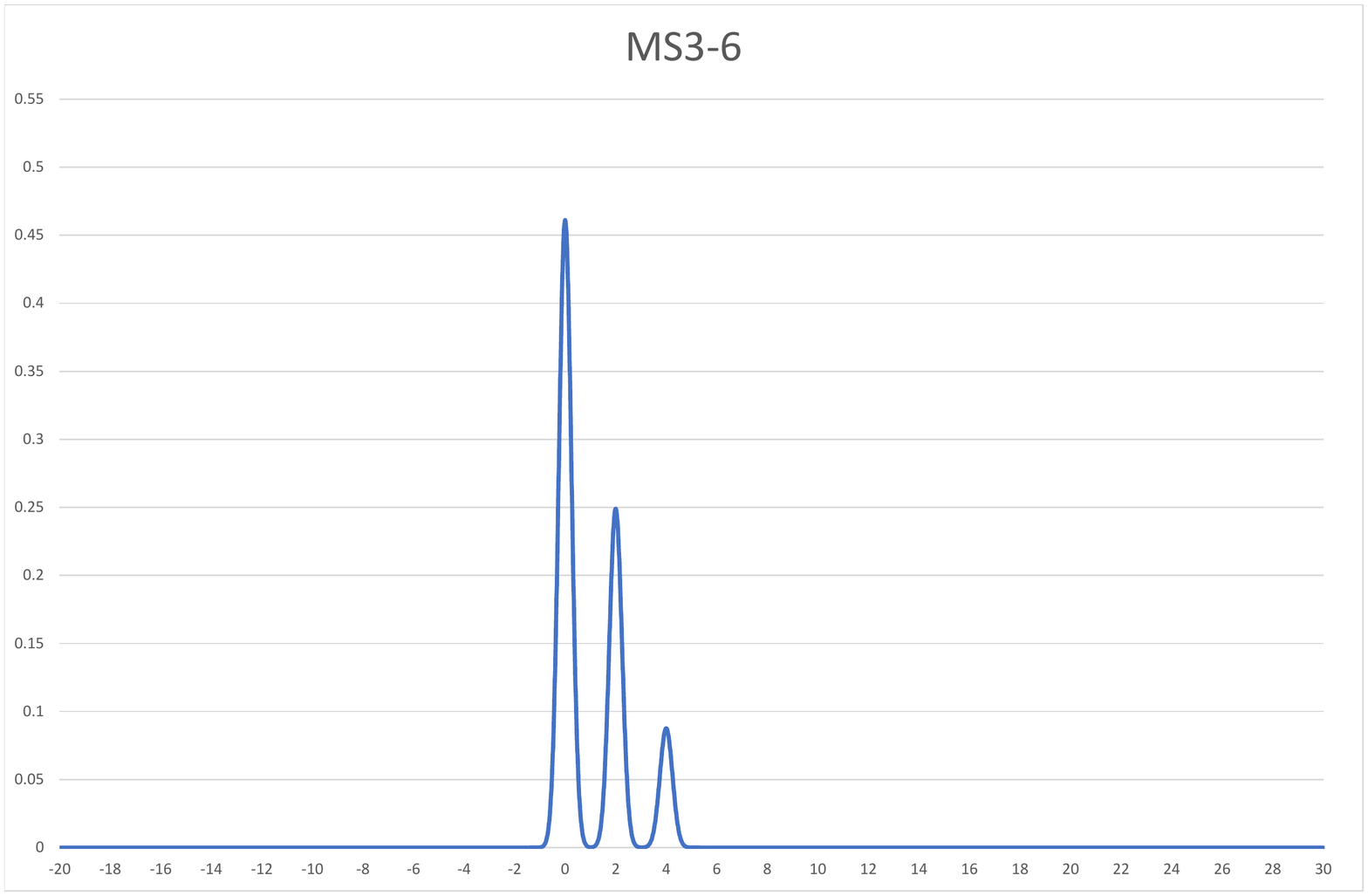} 
	\end{subfigure} 
	\hfill
	\begin{subfigure}[h!]{0.24\textwidth}
		\centering		
		\includegraphics[width=\textwidth]{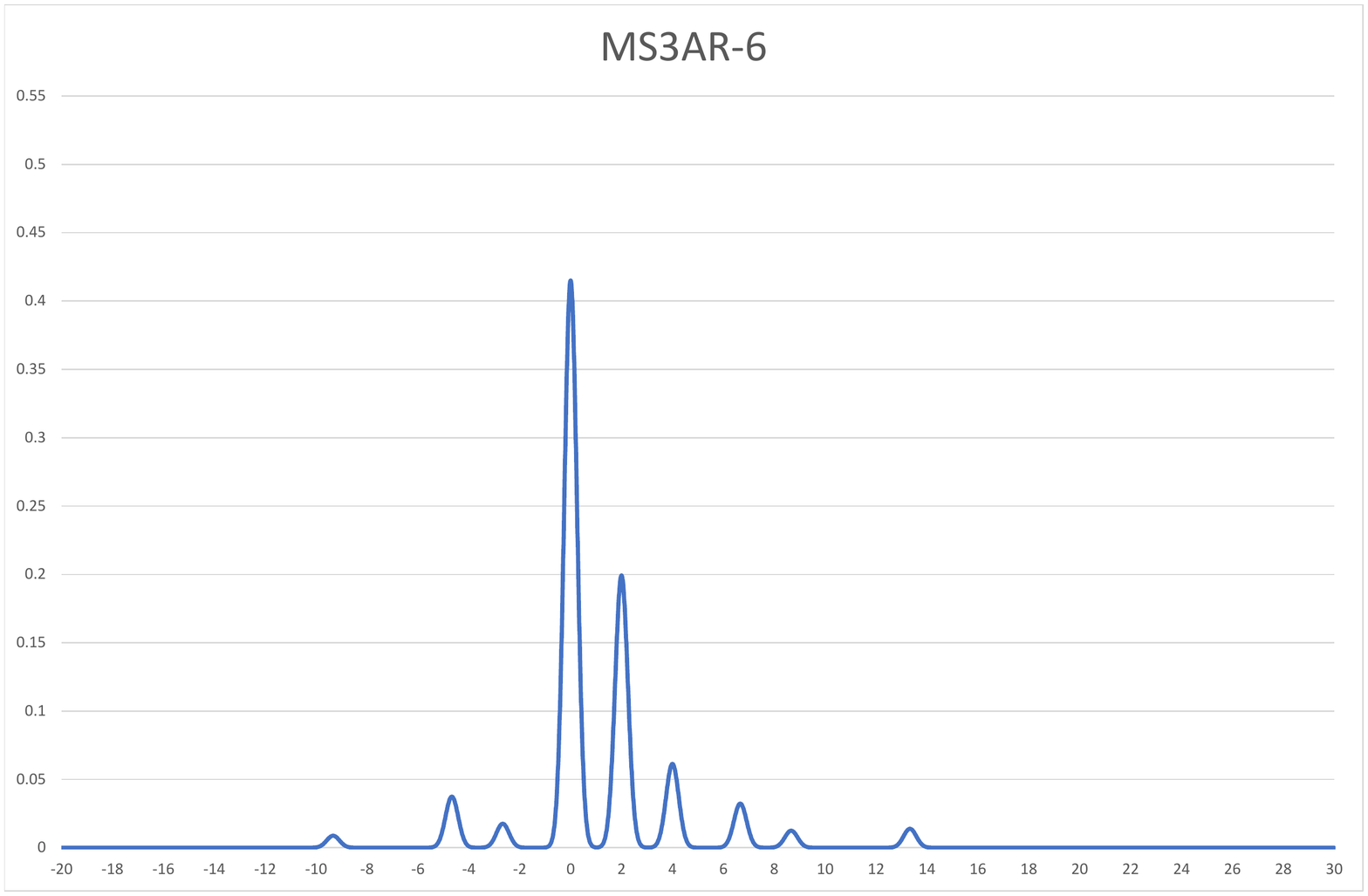} 
	\end{subfigure} 
	\vfill
	\begin{subfigure}[h!]{0.24\textwidth}
		\centering		  
		\includegraphics[width=\textwidth]{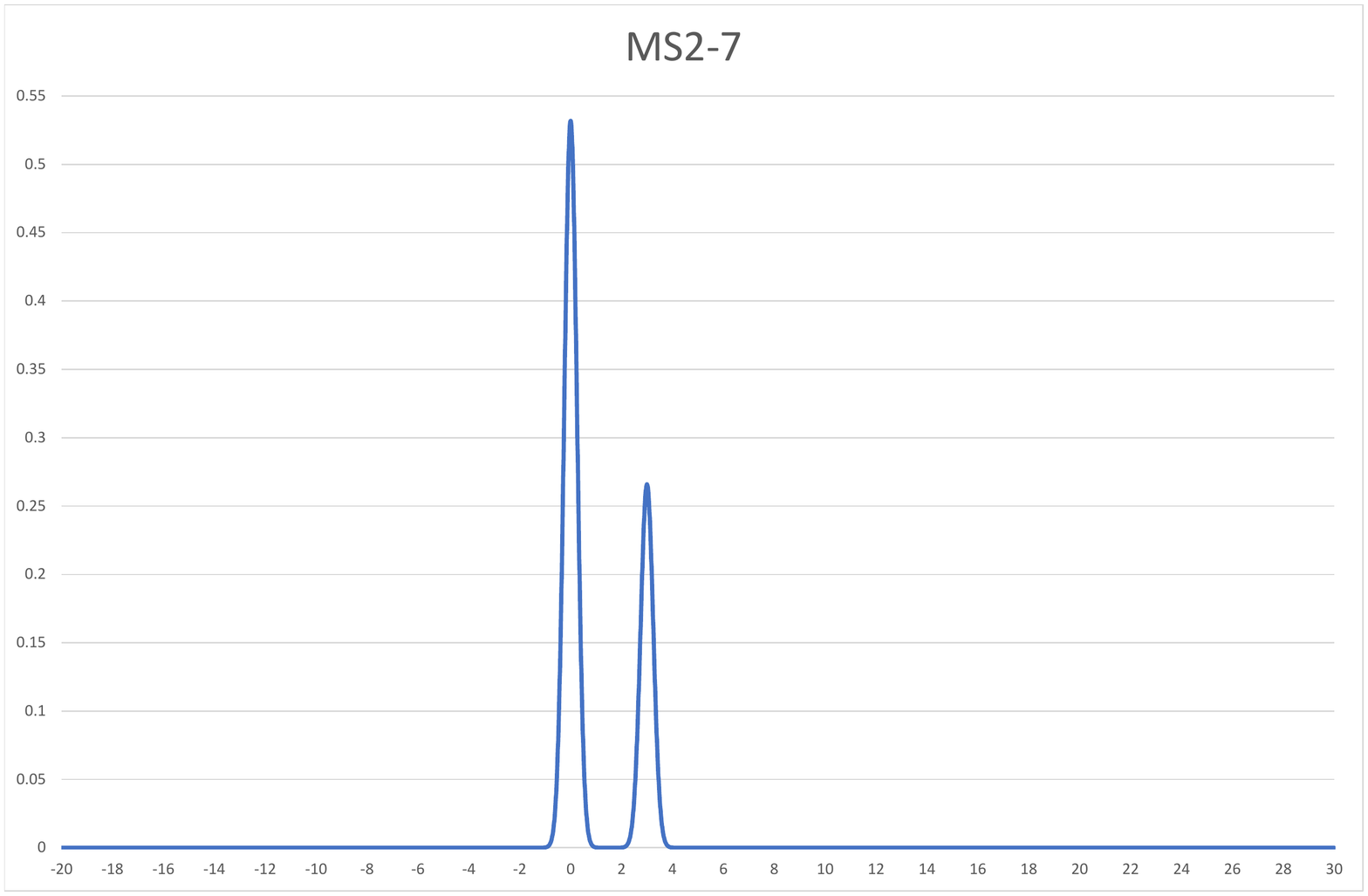}  
	\end{subfigure}
	\hfill 
	\begin{subfigure}[h!]{0.24\textwidth}
		\centering
		\includegraphics[width=\textwidth]{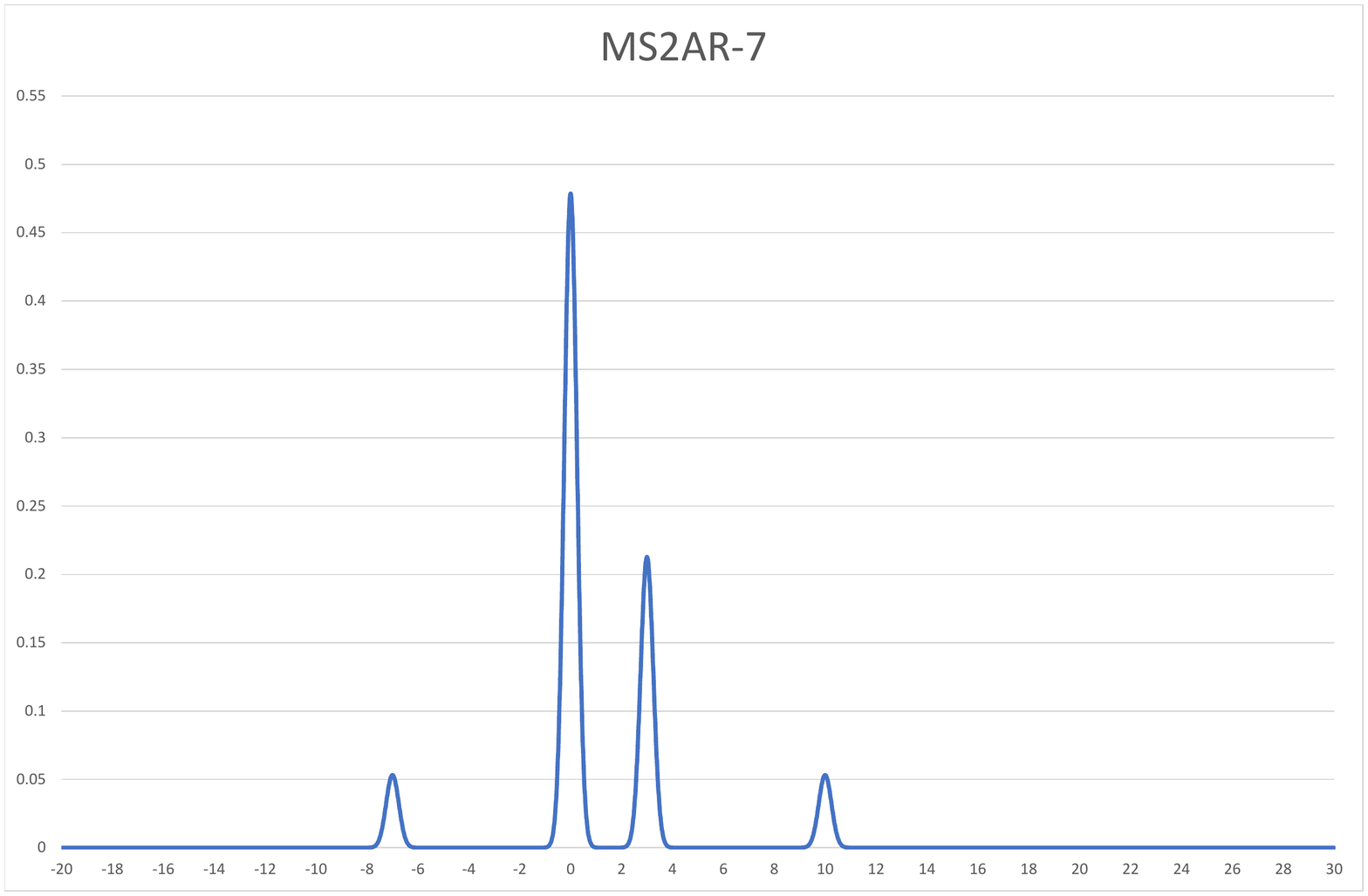}
	\end{subfigure}   
	\hfill
	\begin{subfigure}[h!]{0.24\textwidth}
		\centering		
		\includegraphics[width=\textwidth]{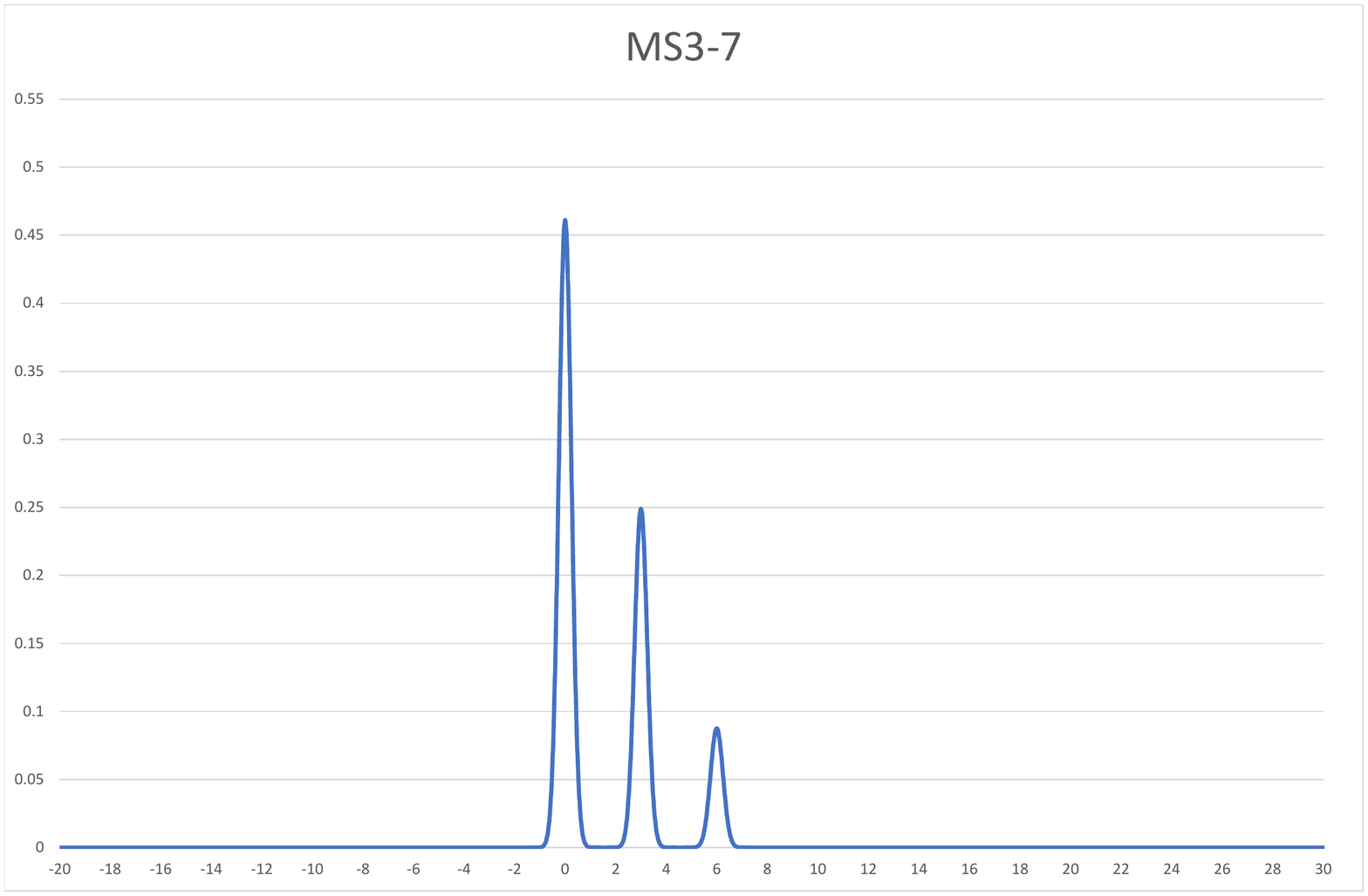} 
	\end{subfigure} 
	\hfill
	\begin{subfigure}[h!]{0.24\textwidth}
		\centering		
		\includegraphics[width=\textwidth]{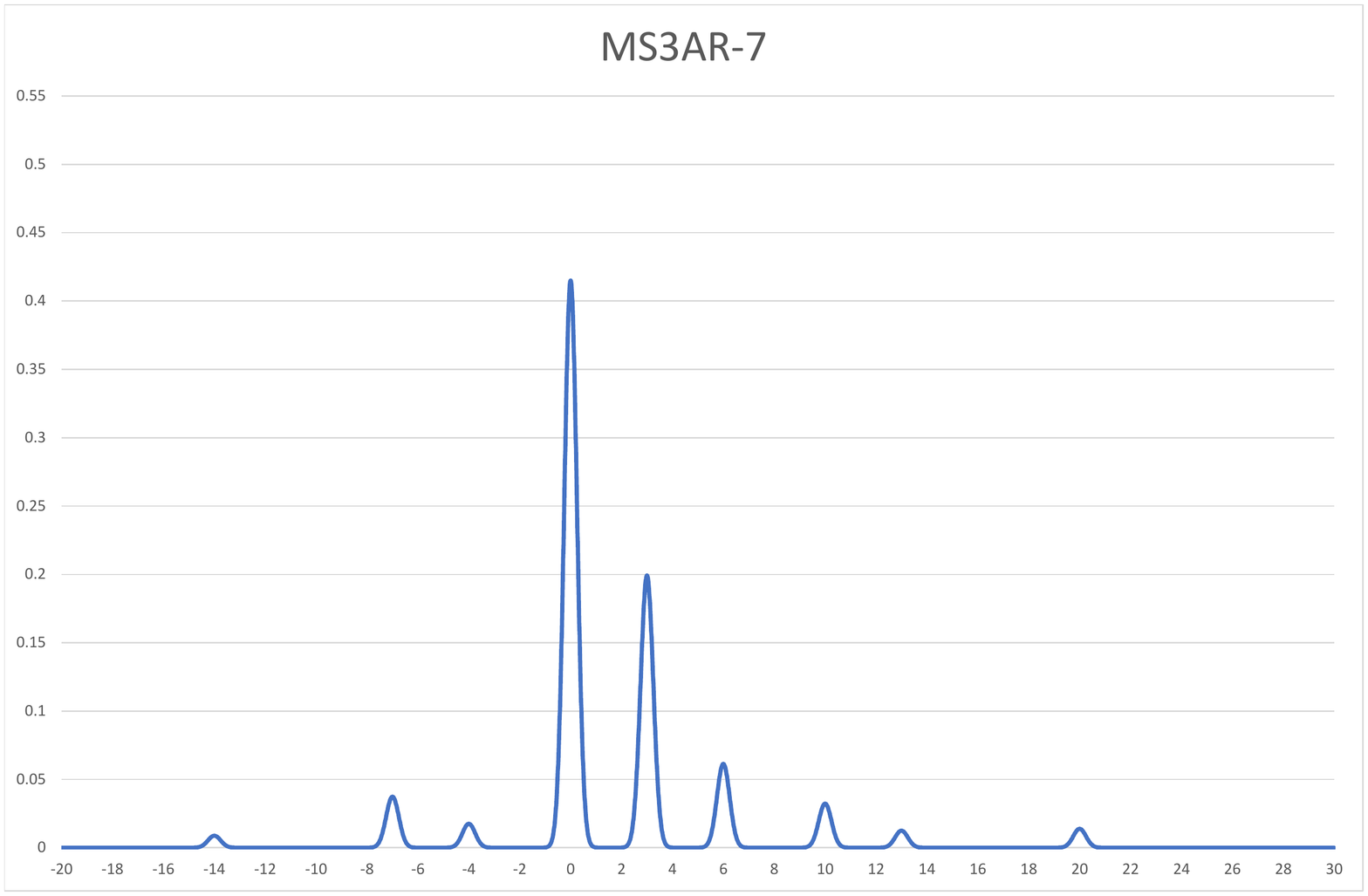} 
	\end{subfigure} 
	\vfill
	\begin{subfigure}[h!]{0.24\textwidth}
		\centering		  
		\includegraphics[width=\textwidth]{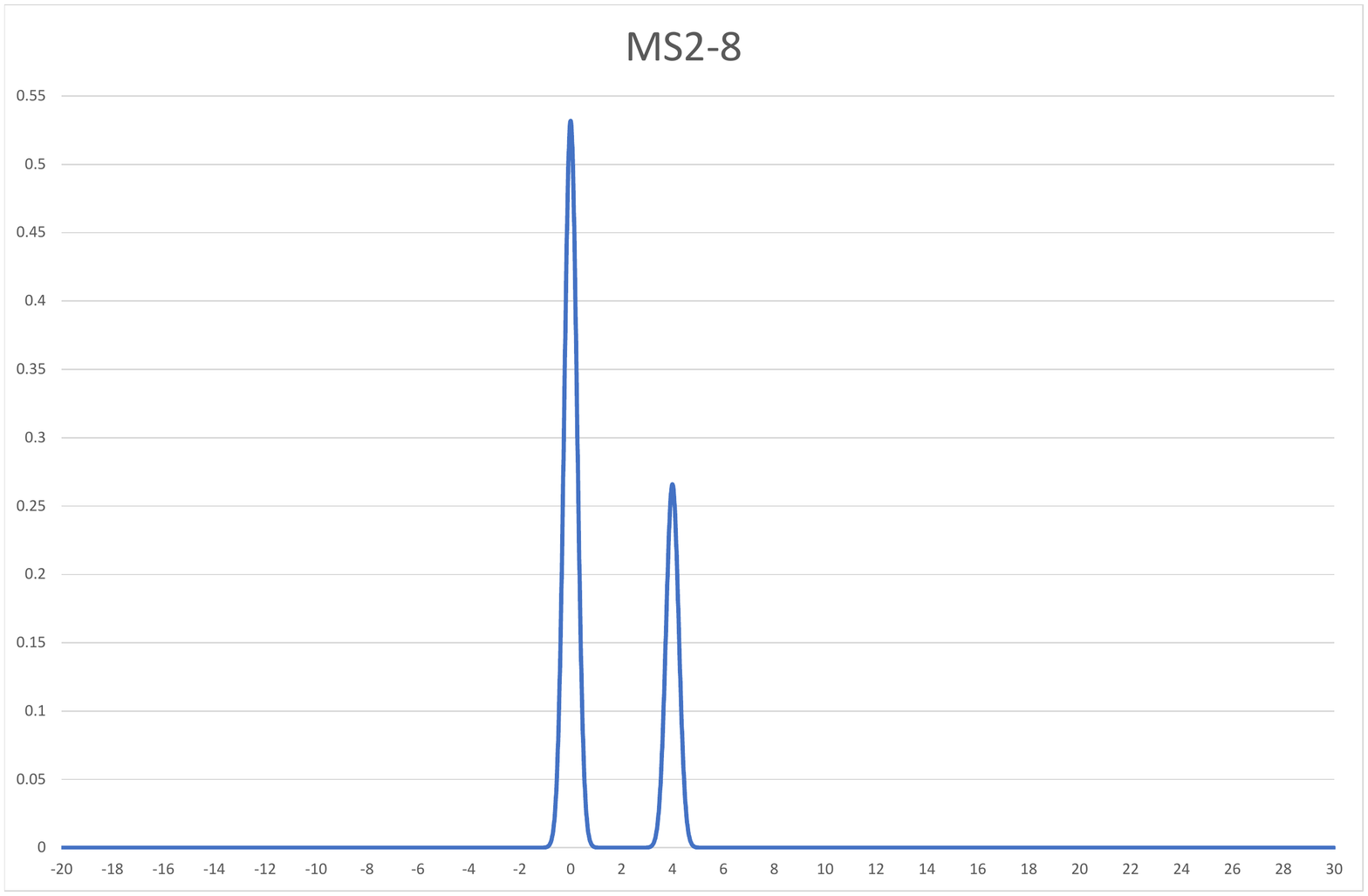}  
	\end{subfigure}
	\hfill 
	\begin{subfigure}[h!]{0.24\textwidth}
		\centering
		\includegraphics[width=\textwidth]{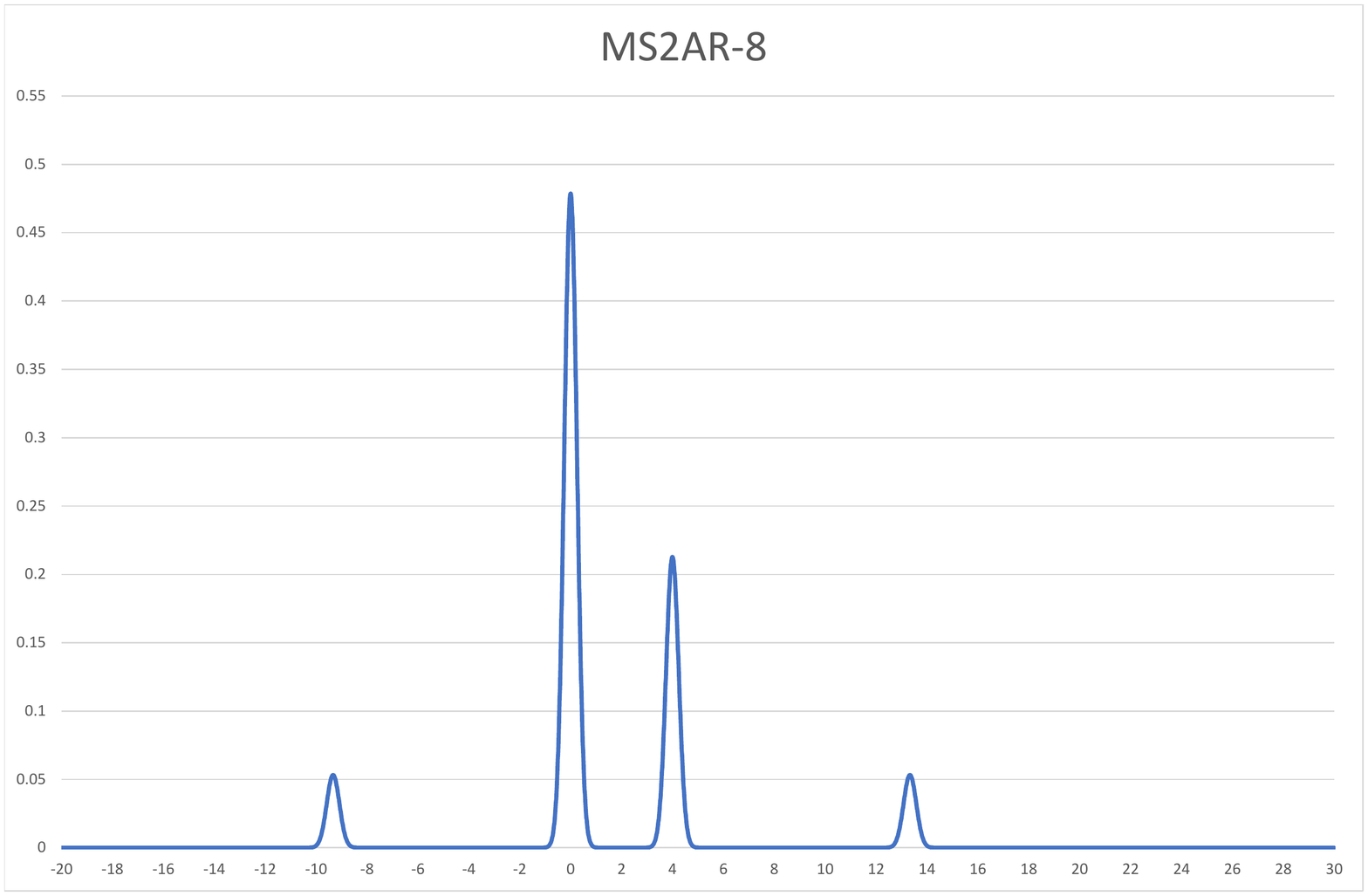}
	\end{subfigure}   
	\hfill
	\begin{subfigure}[h!]{0.24\textwidth}
		\centering		
		\includegraphics[width=\textwidth]{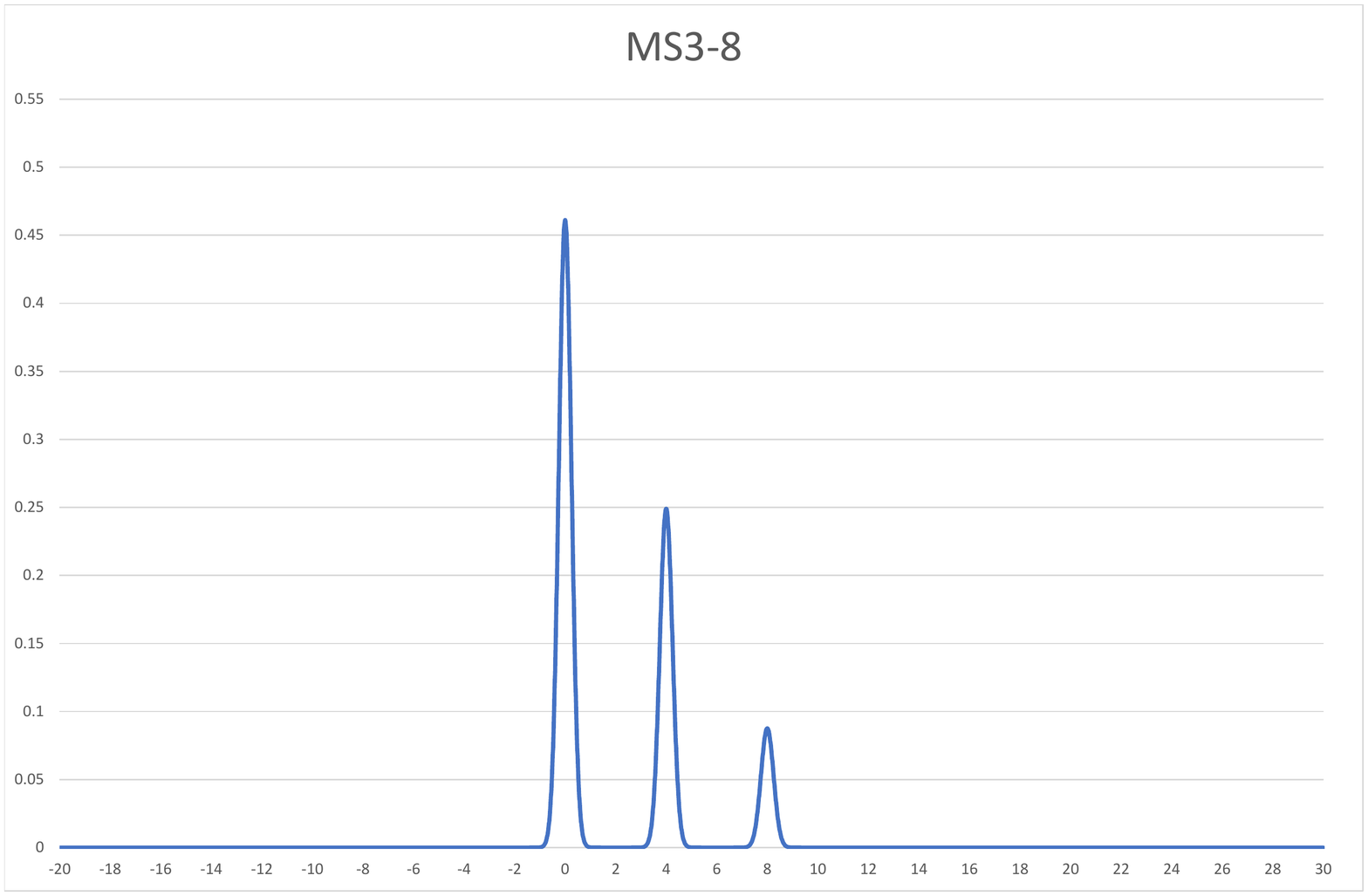} 
	\end{subfigure} 
	\hfill
	\begin{subfigure}[h!]{0.24\textwidth}
		\centering		
		\includegraphics[width=\textwidth]{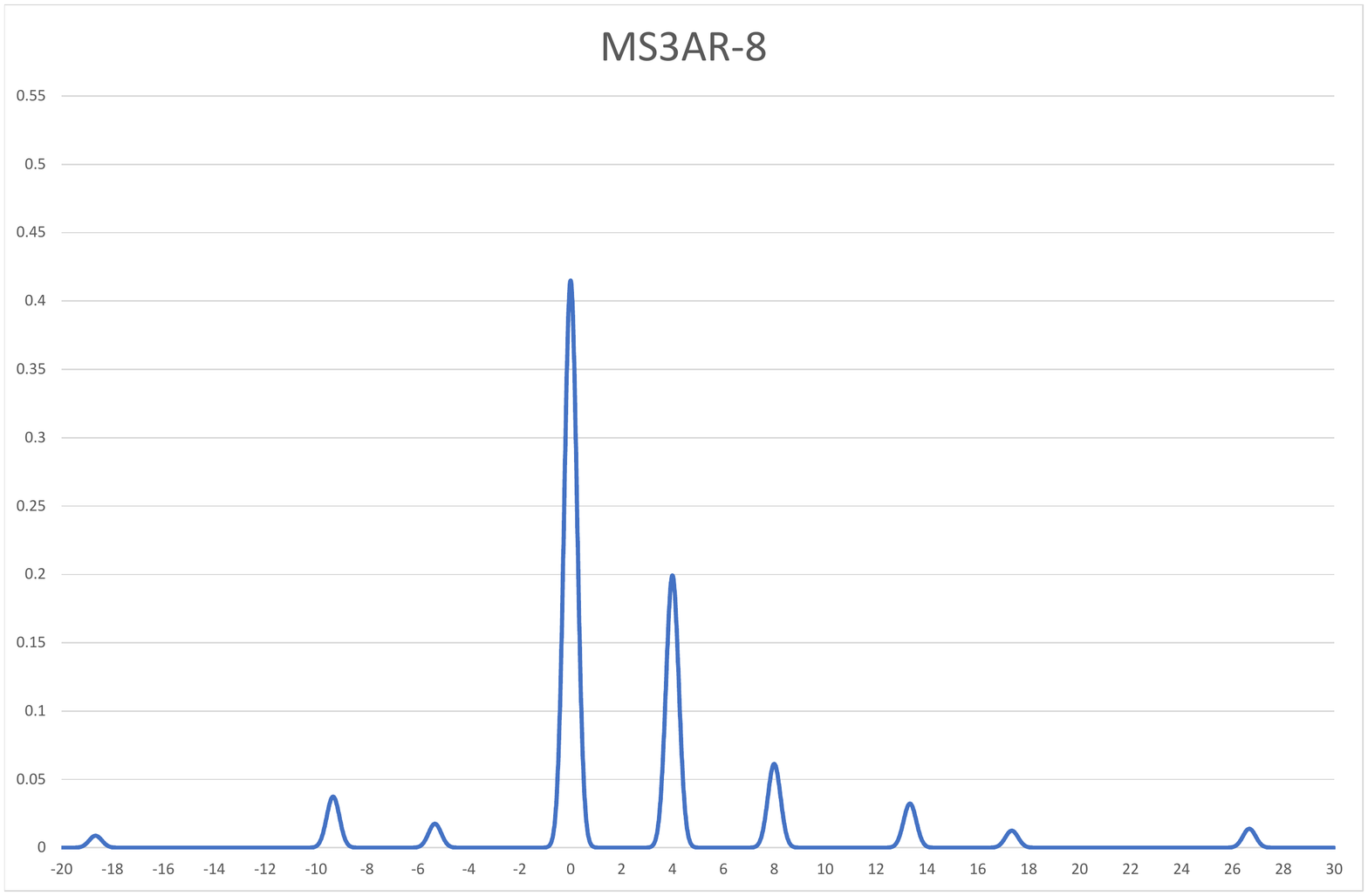} 
	\end{subfigure} 
	\label{fig:ergdist} 
\end{figure}

As the distance between the $\mu_i$ coefficients increases, the existence of  states becomes clearer. This can be better appreciated by looking at Figure \ref{fig:ergdist}, where we show the mixture of Normal distributions obtained using, as mixture weights, the ergodic probability of each state.\footnote{As shown in equation (\ref{density}), each density has  different mixture weights, given by the predicted probabilities $Pr(s_t|I_{t-1})$; the vector of ergodic probabilities is the (normalized) eigenvector associated to the unit eigenvalue of ${\bm P}'$, which can be interpreted as the vector of unconditional probabilities of the state $s_t$ \citep[for details, see][]{Hamilton94}.}  The parameters of the Normal distributions are given by the unconditional means of $y_t$, expressed by $\mu_{s_t}$ in the MS(2) and MS(3) DGPs, and $(\mu_{s_t}-\phi \mu_{s_{t-1}})/(1-\phi)$ in the case of MS(2)--AR(1) and MS(3)--AR(1).\footnote{This implies that, in the case of AR(1) model, the number of states can be seen as $2^k$ \citep[for details, see][]{Hamilton94}.} The number of mixtures is indistinguishable when the $\mu_i$ coefficients are close and the variance is larger (as in MS2--1, MS2AR--1, MS3--1, MS3AR--1); the presence of states is more evident by increasing the distance between $\mu_i$ parameters and decreasing the variance. The presence of the AR parameters increases the number of components of the mixture, but the number of highest peaks are equal to the number of states.

For each DGP we generate 1000 time series of length $T= 100$. For each series we perform two types of analyses: first, considering the number of states known, we verify whether  fuzzy clustering  provides a similar inference on the states obtained from the  estimated MS model; then we verify through the validation indices listed at the end of Section \ref{sec:msfuz} if the clustering algorithm is able to detect the right number of states.  

\begin{table}[H]
\begin{center}
\caption{Data Generating Processes (DGPs) used in Monte Carlo experiments.}\label{tab:mods}
\footnotesize{\begin{tabular}{l c|lc } 
\hline
Label&Parameters&Label&Parameters\\ 
\multicolumn{2}{c|}{DGP: MS(2)}&\multicolumn{2}{c}{DGP: MS(2)--AR(1)}\\
MS2--1&$\mu_1=0$; $\mu_2=1$; $\sigma=0.5$&MS2AR--1&$\mu_1=0$; $\mu_2=1$; $\sigma=0.5$\\
MS2--2&$\mu_1=0$; $\mu_2=2$; $\sigma=0.5$&MS2AR--2&$\mu_1=0$; $\mu_2=2$; $\sigma=0.5$\\
MS2--3&$\mu_1=0$; $\mu_2=3$; $\sigma=0.5$&MS2AR--3&$\mu_1=0$; $\mu_2=3$; $\sigma=0.5$\\
MS2--4&$\mu_1=0$; $\mu_2=4$; $\sigma=0.5$&MS2AR--4&$\mu_1=0$; $\mu_2=4$; $\sigma=0.5$\\
MS2--5&$\mu_1=0$; $\mu_2=1$; $\sigma=0.25$&MS2AR--5&$\mu_1=0$; $\mu_2=1$; $\sigma=0.25$\\
MS2--6&$\mu_1=0$; $\mu_2=2$; $\sigma=0.25$&MS2AR--6&$\mu_1=0$; $\mu_2=2$; $\sigma=0.25$\\
MS2--7&$\mu_1=0$; $\mu_2=3$; $\sigma=0.25$&MS2AR--7&$\mu_1=0$; $\mu_2=3$; $\sigma=0.25$\\				
MS2--8&$\mu_1=0$; $\mu_2=4$; $\sigma=0.25$&MS2AR--8&$\mu_1=0$; $\mu_2=4$; $\sigma=0.25$\\	\hline				
\multicolumn{2}{c|}{DGP: MS(3)}&\multicolumn{2}{c}{DGP: MS(3)--AR(1)}\\
MS3--1&$\mu_1=0$; $\mu_2=1$; $\mu_3=2$; $\sigma=0.5$&MS3AR--1&$\mu_1=0$; $\mu_2=1$; $\mu_3=2$; $\sigma=0.5$\\
MS3--2&$\mu_1=0$; $\mu_2=2$; $\mu_3=4$; $\sigma=0.5$&MS3AR--2&$\mu_1=0$; $\mu_2=2$; $\mu_3=4$; $\sigma=0.5$\\
MS3--3&$\mu_1=0$; $\mu_2=3$; $\mu_3=6$; $\sigma=0.5$&MS3AR--3&$\mu_1=0$; $\mu_2=3$; $\mu_3=6$; $\sigma=0.5$\\
MS3--4&$\mu_1=0$; $\mu_2=4$; $\mu_3=8$; $\sigma=0.5$&MS3AR--4&$\mu_1=0$; $\mu_2=4$; $\mu_3=8$; $\sigma=0.5$\\
MS3--5&$\mu_1=0$; $\mu_2=1$; $\mu_3=2$; $\sigma=0.25$&MS3AR--5&$\mu_1=0$; $\mu_2=1$; $\mu_3=2$; $\sigma=0.25$\\
MS3--6&$\mu_1=0$; $\mu_2=1$; $\mu_3=2$; $\sigma=0.25$&MS3AR--6&$\mu_1=0$; $\mu_2=2$; $\mu_3=4$; $\sigma=0.25$\\
MS3--7&$\mu_1=0$; $\mu_2=1$; $\mu_3=2$; $\sigma=0.25$&MS3AR--7&$\mu_1=0$; $\mu_2=3$; $\mu_3=6$; $\sigma=0.25$\\
MS3--8&$\mu_1=0$; $\mu_2=1$; $\mu_3=2$; $\sigma=0.25$&MS3AR--8&$\mu_1=0$; $\mu_2=4$; $\mu_3=8$; $\sigma=0.25$\\
\hline
\end{tabular}}
\end{center}
\begin{tablenotes}[flushleft]
\footnotesize 
\item The AR(1) coefficient, when present, is equal to 0.7 in all DGPs. The transition probabilities in the MS(2) and MS(2)--AR(1) DGPs are $p_{11}=0.9$ and $p_{22}=0.8$; in the MS(3) and MS(3)--AR(1) DGPs are  $p_{11}=0.9$, $p_{12}=0.07$, $p_{22}=0.8$,  $p_{21}=0.15$, $p_{33}=0.7$, $p_{32}=0.2$. 
\end{tablenotes}
\end{table}

\subsection{Inference on the state}\label{sec:inf}
As said, the inference on the state of MS models is performed by assigning the observation at time $t$ to the state with the highest smoothed probability. In our first experiment we generate data from the 32 DGPs shown in Table \ref{tab:mods},  estimate the corresponding MS model, derive the smoothed probabilities, assign each observation to a state, and finally compare the clustering to the true one using the Rand index \citep{ran71}:
\begin{equation}
R= 
\left\{\binom{T}{2}-1/2\left[\sum_{i=1}^k\left(\sum_{j=1}^k {T_{ij}}\right)^2+\sum_{j=1}^k\left(\sum_{i=1}^k {T_{ij}}\right)^2+\sum_{i=1}^k\sum_{j=1}^kT_{ij}^2\right]\right\}/\binom{T}{2}
\label{rand}
\end{equation}
where $T_{ij}$ is the number of series
belonging to the group $i$ in the true clustering and assigned to the
group $j$ in the MS clustering. The Rand index $R$ ranges in
$[0,1]$, with maximum value in the case of perfect match between true and MS clustering, while it is 0 when there are no similarities among them. We calculated $R$ for each simulated time series of each DGP; in the first columns of  Table \ref{tab:rand} we show the five--number summary (minimum, first quartile, median, third quartile, maximum)\footnote{We prefer to show these values in a table and not in  box--plots because the latter is difficult to be visualized when the variability of results is close to zero.}  of the empirical distribution of $R$ for each DGP. 

It is interesting to note that, despite the fact that the estimated model is correctly specified, when the means are close to each other--see also Figure \ref{fig:ergdist}-- it is not trivial to obtain the correct classification of the observations (models MS2--1 , MS3--1), and the problem is more evident when there is an autoregressive term (MS2AR--1, MS2AR--2, MS3AR--1, MS3AR--2). The Rand index increases when the variance of each state is smaller (models labeled with numbers from 4 to 8). However, excluding the MS3AR--1 and the MS2AR--1 cases, in each DGP more than 50\% of the simulations provide a Rand index greater than 0.8. The 100\% correct classifications are achieved when the distance among the means of each state is larger and the variance is smaller (models MS3--7, MS3--8, MS2AR--7, MS2AR--8, MS3AR--7, MS3AR--8).

Considering the fuzzy classification, obtained using the squared Euclidean distance in (\ref{f_cen}), and comparing this classification with the true one (second block of columns in Table \ref{tab:rand}), we note a certain difference with respect to the previous comparison only in the minimum value of the Rand index and a better behavior for MS2--7 and MS2--8. In practice, the classification obtained in a nonparametric way and not considering the model that generates the data, shows very similar results to the case using the right model (not known in practical cases), which requires the estimation step. This is confirmed by observing the third block of columns of Table \ref{tab:rand}, where the MS and the fuzzy approach are compared (in equation (\ref{rand}) the index $i$ now refers to the MS classification and $j$ to the fuzzy classification); classification differences are reduced when the distance between means is larger and the variability is smaller.

\begin{table}[H]
	\begin{center}
		\caption{Summary Statistics of Rand Index for MS Inference on the state and Fuzzy k-means clustering in 1000 Monte Carlo experiments.}\label{tab:rand}{\tiny }
		  \resizebox{1\textwidth}{!}{
		  \begin{tabular}{l|ccccc|ccccc|ccccc|} 
				\hline
				{DGP}&\multicolumn{15}{c|}{{Rand Index}}\\ 
				& \multicolumn{5}{c|}{{MS-True State}} & \multicolumn{5}{c|}{{Fuzzy-True State}}& \multicolumn{5}{c|}{{Fuzzy-MS}} \\
				& {Min.} & {Q1} & {Q2} & {Q3} & {Max.} & {Min.} & {Q1} & {Q2} & {Q3} & {Max.} & {Min.} & {Q1} & {Q2} & {Q3} & {Max.} \\
				{MS2--1}& {0.50} &{0.76}  &{0.82}   &{0.87}&  {1.00} &{0.50}	&{0.64}	&{0.69}&   {0.74}  &{0.89}	&{0.50}	&{0.68}&   {0.74}&{0.80}  &	{1.00} \\
				{MS2--2  }&0.79&	0.94&	0.96&	0.98&	1.00&	0.52&	0.92&	0.96&	0.98&	1.00&	0.54&	0.92&	0.94&	0.96&	1.00
				
				  \\
				{MS2--3  }&0.94&	0.98&	0.98&	0.98&	0.98&	0.94	&1.00&	1.00&	1.00&	1.00&	0.92	&0.98&	0.98&	0.98&	0.98
				\\
				{MS2--4  }&0.92&	0.98&	0.98&	0.98&	0.98&	0.74&	1.00&	1.00&	1.00&	1.00&	0.80&	0.98&	0.98&	0.98&	0.98
				\\
				{MS2--5  }&0.82&	0.97&	0.98&	1.00&	1.00&	0.52&	0.90&	0.94&	0.97&	1.00&	0.50&	0.90&	0.95&	0.98&	1.00
				\\
				{MS2--6  }&0.98&0.98&	0.98&0.98&	0.98&	0.55&	1.00&	1.00&	1.00&	1.00&	0.54&	0.98&	0.98&	0.98&	0.98
				\\
				{MS2--7  }&0.98	&0.98&0.98&0.98	&0.98&	1.00&	1.00&1.00	&1.00&1.00&0.98&0.98&0.98&0.98&0.98
				\\
				{MS2--8  }&0.98	&0.98&0.98	&0.98&	0.98&	1.00&	1.00&	1.00&	1.00&	1.00&	0.98&	0.98&	0.98&	0.98&	0.98
				\\ 				
				{MS2AR--1}&0.49	&0.56&	0.64&	0.73&	0.98&	0.49&	0.56&	0.60&	0.66&	0.90&	0.49&	0.54&	0.66&	0.79&	0.96
				\\
				{MS2AR--2}&0.52&	0.83&	0.90&	0.96&	1.00&	0.50&	0.79&	0.85&	0.90&	1.00&	0.49&	0.76&	0.85&	0.92&	1.00
				\\
				{MS2AR--3}&0.63&	0.98&	1.00&	1.00&	1.00&	0.57&	0.94&	0.98&	1.00&	1.00&	0.58&	0.94&	0.98&	1.00&	1.00
				\\
				{MS2AR--4}&0.82&	1.00&	1.00&	1.00&	1.00&	0.51&	1.00&	1.00&	1.00&	1.00&	0.51&	1.00&	1.00&	1.00&	1.00
				\\
				{MS2AR--5}&0.51&	0.82&	0.90&	0.96&	1.00&	0.50&	0.78&	0.85&	0.90&	1.00&	0.50&	0.76&	0.85&	0.90&	1.00
				\\
				{MS2AR--6}&0.89&	1.00&	1.00&	1.00&	1.00&	0.50&	1.00&	1.00&	1.00&	1.00&	0.50&	1.00&	1.00&	1.00&	1.00
				\\
				{MS2AR--7}&1.00&	1.00&	1.00&	1.00&	1.00&	0.98&	1.00&	1.00&	1.00&	1.00&	0.98&	1.00&	1.00&	1.00&	1.00
				\\				
				{MS2AR--8}&1.00&	1.00&	1.00&	1.00&	1.00&	1.00&	1.00&	1.00&	1.00&	1.00&	1.00&	1.00&	1.00&	1.00&	1.00
				\\	
			    {MS3--1}&0.53&	0.79&	0.84&	0.89&	0.98&	0.47&	0.63&	0.67&	0.71&	0.88&	0.48&	0.65&	0.69&	0.76&	0.98
			    \\
				{MS3--2}&0.85&	0.97&	0.98&	1.00&	1.00&	0.58&	0.90&	0.94&	0.96&	1.00&	0.56&	0.91&	0.95&	0.97&	1.00
				\\
				{MS3--3}&0.96&	1.00&	1.00&	1.00&	1.00&	0.56&	0.99&	1.00&	1.00&	1.00&	0.56&	1.00&	1.00&	1.00&	1.00
				\\
				{MS3--4}&0.98&	1.00&	1.00&	1.00&	1.00&	0.57&	1.00&	1.00&	1.00&	1.00&	0.57&	1.00&	1.00&	1.00&	1.00
				\\
				{MS3--5}&0.82&	0.97&	0.98&	1.00&	1.00&	0.52&	0.90&	0.94&	0.97&	1.00&	0.50&	0.90&	0.95&	0.98&	1.00
				\\
				{MS3--6}&0.98&	1.00&	1.00&	1.00&	1.00&	0.56&	1.00&	1.00&	1.00&	1.00&	0.56&	1.00&	1.00&	1.00&	1.00
				\\
				{MS3--7}&1.00&	1.00&	1.00&	1.00&	1.00&	0.57&	1.00&	1.00&	1.00&	1.00&	0.57&	1.00&	1.00&	1.00& 1.00
				\\
				{MS3--8}&1.00&	1.00&	1.00&	1.00&	1.00&	0.62&	1.00&	1.00&	1.00&	1.00&	0.62&	1.00&	1.00&	1.00&	1.00
				\\ 
				{MS3AR--1}&0.38&	0.58&	0.64&	0.71&	0.96&	0.41&	0.58&	0.62&	0.66&	0.88&	0.37&	0.61&	0.67&	0.74&	0.94
				\\
				{MS3AR--2}&0.50&	0.81&	0.89&	0.95&	1.00&	0.44&	0.70&	0.78&	0.86&	1.00&	0.39&	0.69&	0.77&	0.87&	0.99
				\\
				{MS3AR--3}&0.62&	0.97&	1.00&	1.00&	1.00&	0.52&	0.92&	0.97&	0.99&	1.00&	0.52&	0.91&	0.96&	0.98&	1.00
				\\
				{MS3AR--4}&0.68	&1.00&	1.00&	1.00&	1.00&	0.55&	0.99&	1.00&	1.00&	1.00&	0.55&	0.99&	1.00&	1.00&	1.00
				\\
				{MS3AR--5}&0.52	&0.81&	0.89&	0.95&	1.00&	0.50	&0.71&	0.80&	0.87&	1.00&	0.43&	0.70&	0.78&	0.88&	1.00
				\\
				{MS3AR--6}&0.70	&1.00&	1.00&	1.00&	1.00&	0.53&	0.99&	1.00&	1.00&	1.00&	0.53&	0.99&	1.00&	1.00&	1.00
				\\
				{MS3AR--7}&1.00	&1.00&	1.00&	1.00&	1.00&	0.55&	1.00&	1.00&	1.00&	1.00&	0.55&	1.00&	1.00&	1.00&	1.00
				
				\\
				{MS3AR--8}&1.00&	1.00&	1.00&	1.00&	1.00&	0.63&	1.00&	1.00&	1.00&	1.00&	0.63&	1.00&	1.00&	1.00&	1.00
				\\
				\hline
		\end{tabular}}
	\end{center}
\begin{tablenotes}[flushleft]
\footnotesize 
\item The summary statistics are the minimum (Min), the first quartile (Q1), the median (Q2), the third quartile (Q3), the maximum (Max) of the Rand index for each set of Monte Carlo experiments. MS refers to the classification obtained by the MS inference on the state, Fuzzy to the classification obtained by the Fuzzy k--means procedure of clustering, True State is the correct  classification of data. 
\end{tablenotes}
\end{table}
\newpage
\subsection{Identification of the number of states}\label{sec:id}
Having empirical evidence of consistent classifications between the typical  parametric MS procedure and the fuzzy nonparametric clustering, the next step is to verify if it is possible to use the usual criteria for detecting the number of clusters to identify the number of states in MS models. We use the same datasets generated for the experiment illustrated in subsection \ref{sec:inf} to apply the criteria for the detection of the number of clusters based on the indices PC, PE, MPC, ASW, ASWF, XB. In Table \ref{tab:ind} we show the percentage of correct identification of the number of states for each index.

Considering DGPs with 2 states, in general all indices are able to detect the correct number of states when the differences between the two means is greater than 1 whatever the variance,  that is excluding the cases MS2--1, MS2--5, MS2AR--1, MS2AR--5. PE shows a good performance also in the last four cases, with a  success rate between 60.1\% (MS2--5) and 90.7\%  (MS2AR-5); second best seems to be the PC index. The indices are often able to detect the correct number of states when the differences between means are larger (DGPs with suffix 3, 4, 7 and 8), equal or very close to 100\%. 

The cases with 3 states have more uncertainty in detecting  the correct number of groups for small differences in means (MS3--1, MS3--2, MS3AR--1, MS3AR--2), but in this case the lower variance seems to promote better performance.  The presence of the autoregressive term in DGP seems to worsen the success rate.

In general, the PE index shows the best performance for the 2--state case, presenting the highest percentage of correct detection in the 16 corresponding DGPs  (in 6 cases 100\%). In the 3--state case there is no a clear preferred index: XB and MPC seem better in half of cases, but the differences in success rate are very small. By summing the success rate in each column of Table \ref{tab:ind}, the PE index shows the highest score with ASW being second best.

\begin{figure}[H]
	\caption{Empirical distributions of the identified number of states with six indices in 1000 Monte Carlo experiments and 4 DGPs.}
\begin{tabular}{cc}
\includegraphics[width=0.5\linewidth,height=0.35\textwidth]{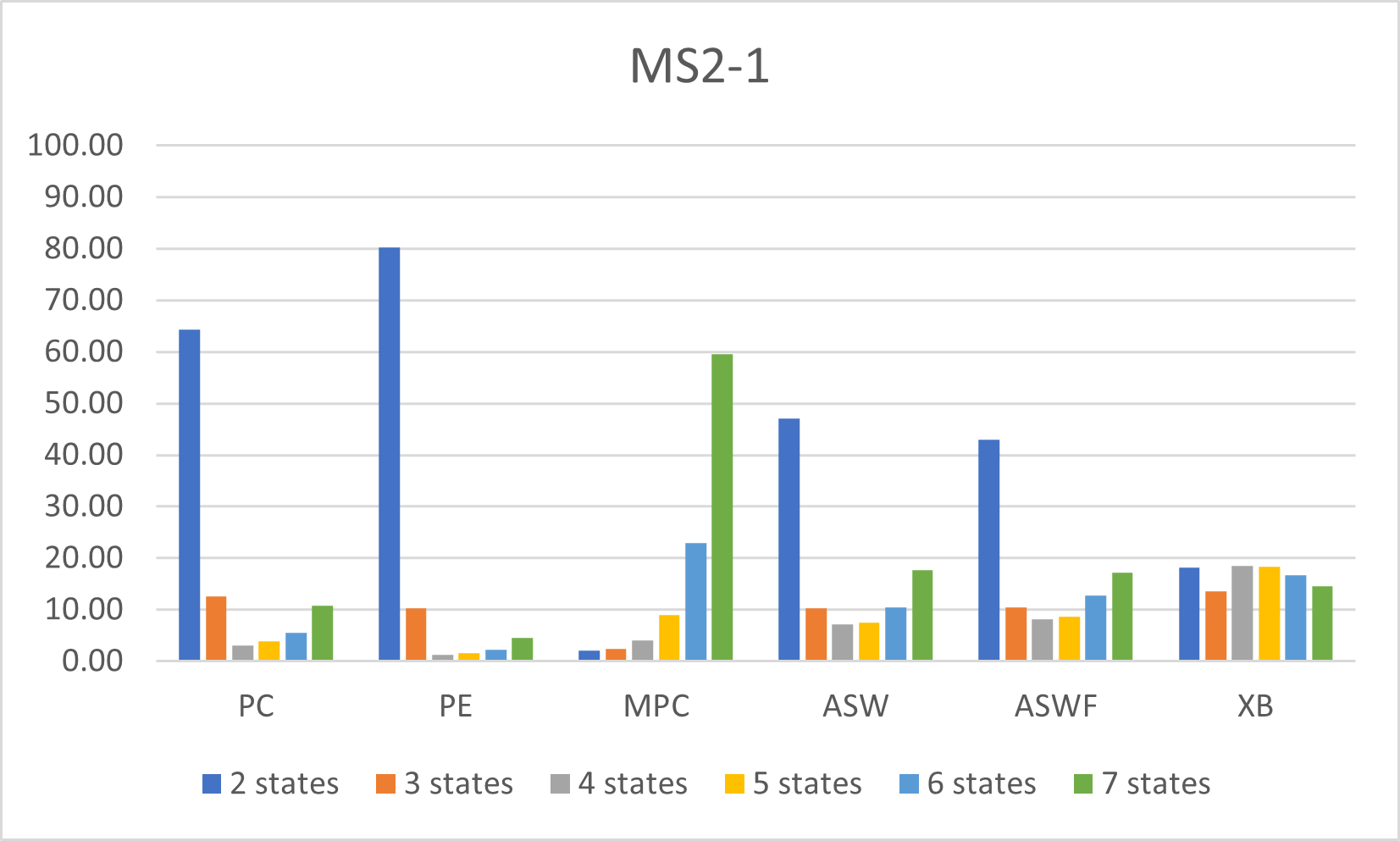}~\includegraphics[width=0.5\linewidth,height=0.35\textwidth]{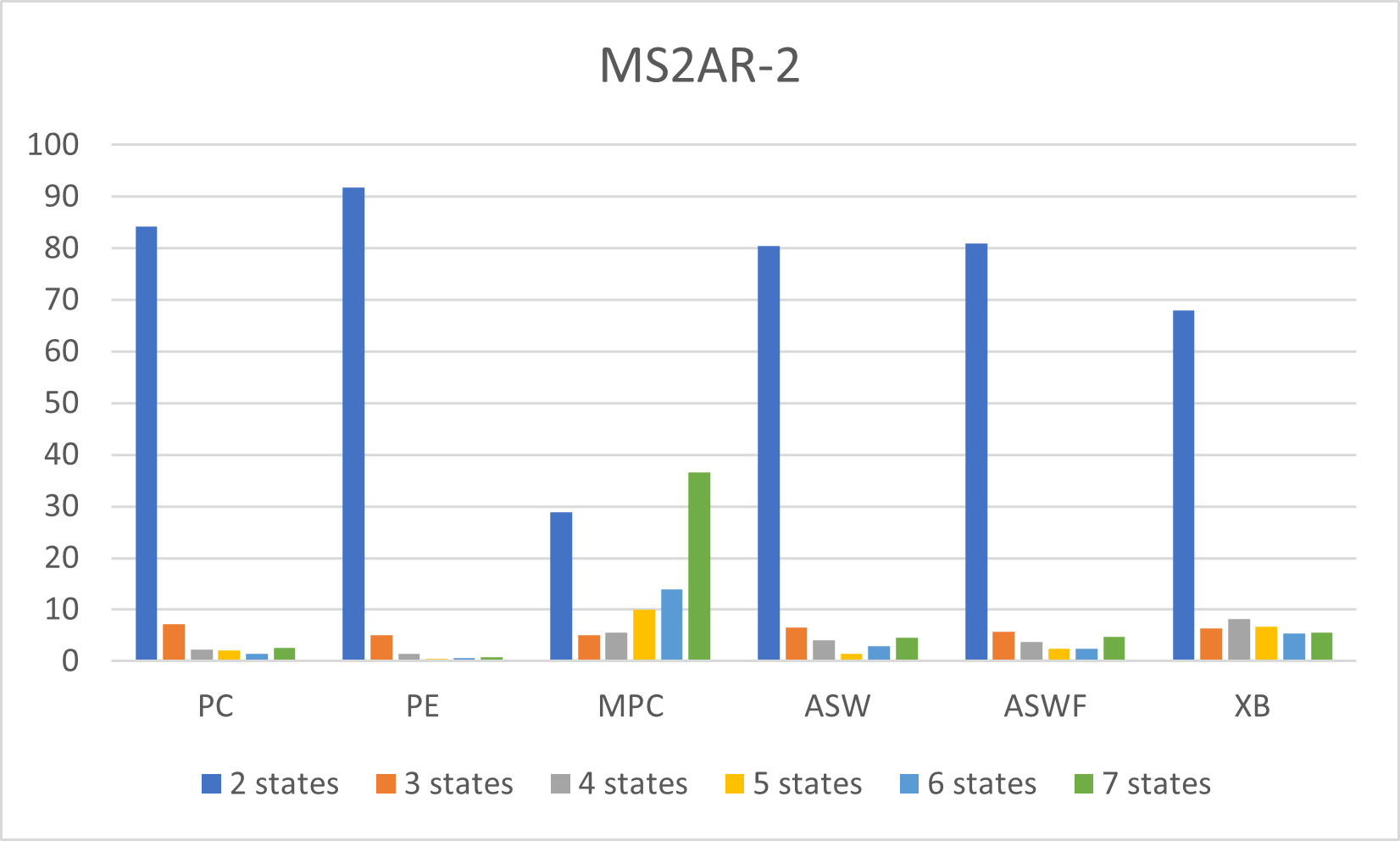}\\
\includegraphics[width=0.5\linewidth,height=0.35\textwidth]{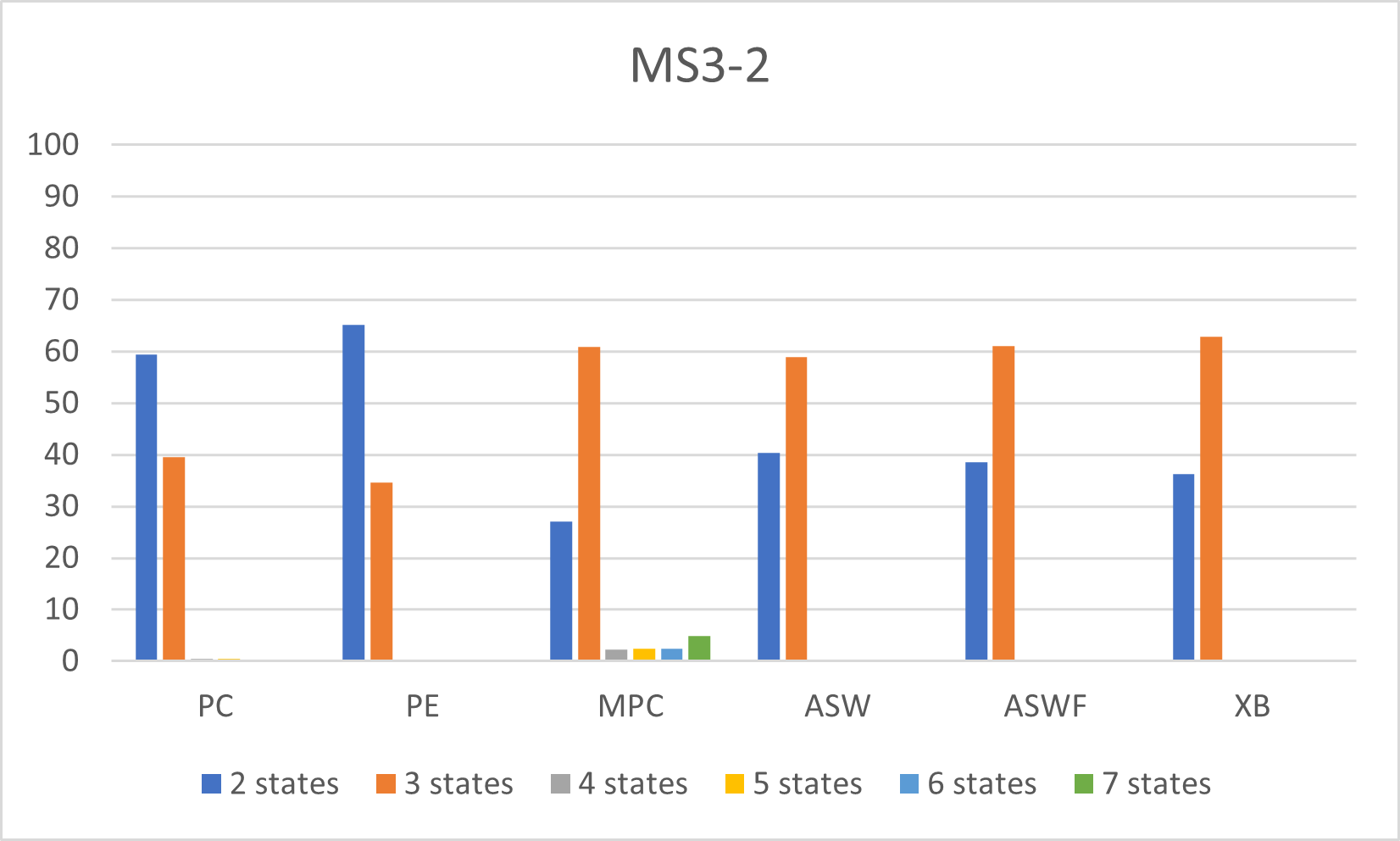}~\includegraphics[width=0.5\linewidth,height=0.35\textwidth]{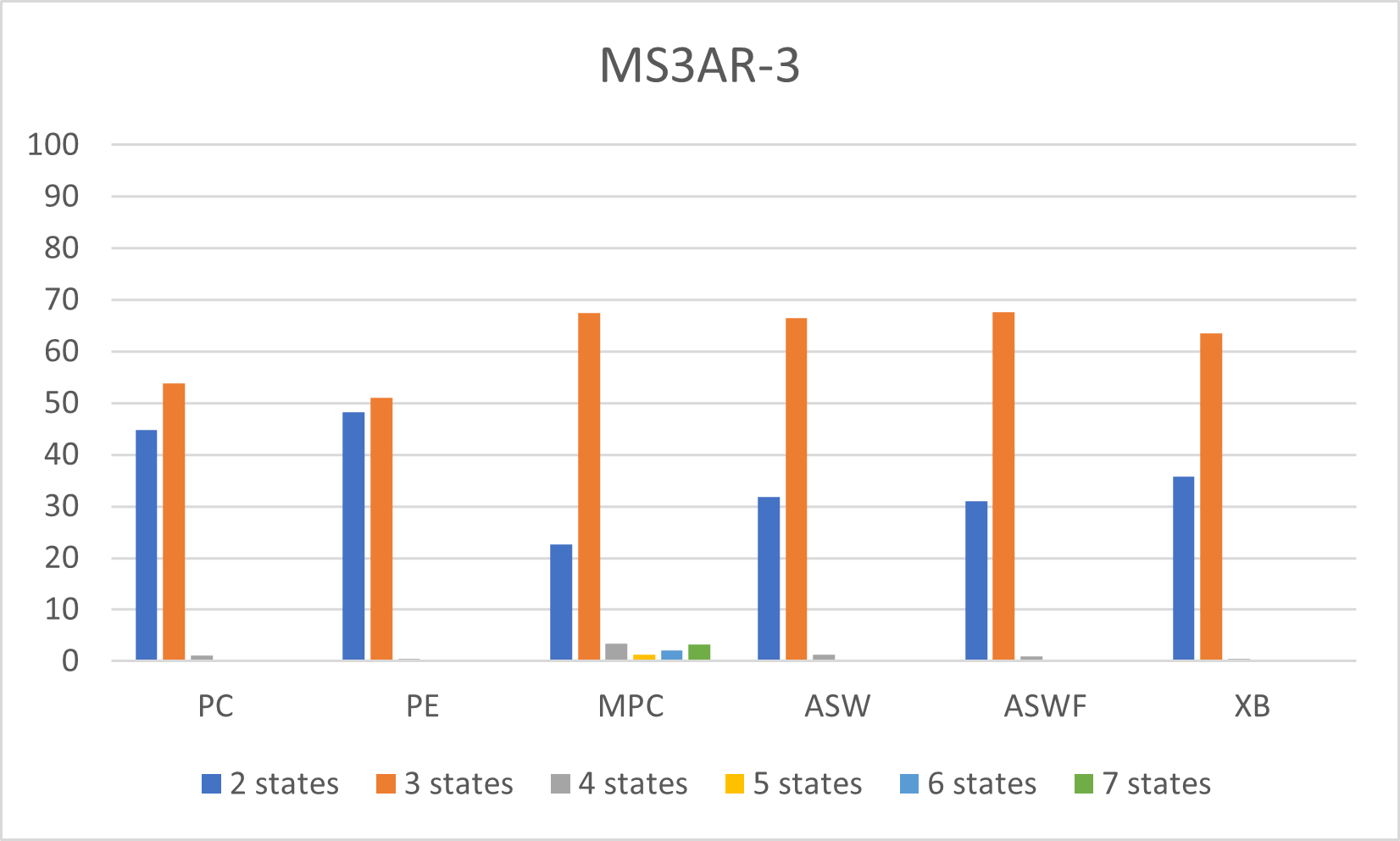} 
\end{tabular} \label{fig:ist}
\end{figure}

In order to have a support in devising a strategy to identify the number of regimes, we have analyzed in more detail the empirical distribution of the identified number of states with each of the six indices. In Figure \ref{fig:ist} we compare these distributions relating to 4 DGPs, where the identification of the number of states seems a bit puzzling.\footnote{Graphs relative to all other DGPs are available upon request.} In the 2-state case (top graphs) PE and PC perform clearly better than other indices, with a clear mode on $k=2$; MPC favors a higher number of states, and XB has uniform behavior in the MS2--1 case. In the 3-state cases, PC and PE are bi--modal distributions (favoring both states 2 and 3), while the performance of the other 4 indices is better, with higher probability on $k=3$; MPC still shows greater variability than the other indices.

Our conclusion is that, when the distance between the mean levels of states is not large (say less than 2), it is convenient to compute PC, PE, ASW and ASWF and check whether they support consistent results in terms of identifying the number of states. When the distance between levels is greater than 2, the six indices appear to provide the same (usually correct) solutions. After the estimation, a comparison between MS inference and fuzzy clustering with the Rand index is a good practice to have some sort of validation of the identified number of states.

\begin{table}[H]
\begin{center}
\caption{Identification of the number of states in 1000 Monte Carlo experiments and 32 Data Generating Processes (DGPs) with several indices}\label{tab:ind}
\footnotesize{\begin{tabular}{l | rrrrrr } 
\hline
&\multicolumn{6}{c}{Indices}\\
DGP&PC&PE&MPC&	ASW &	ASWF&XB\\ \hline
MS2--1&64.3&	80.2&	2.1&	47.0&	42.9&	18.2\\
MS2--2&99.2&	99.6&	86.1&	99.2&	99.3&	98.9\\
MS2--3&100.0&100.0&100.0&100.0&100.0&100.0\\
MS2--4&100.0&100.0&99.9&99.9&99.9&99.9\\
MS2--5&54.7&	60.1&	26.0&	39.7&	40.7&	38.4\\
MS2--6&99.9&99.9&99.9&99.9&99.9&99.9\\
MS2--7&100.0&100.0&100.0&100.0&100.0&100.0\\
MS2--8&100.0&100.0&100.0&100.0&100.0&100.0\\
MS2AR--1&58.4&	74.0&	1.6&	37.6&	34.1&	13.7\\
MS2AR--2&84.3&	91.8&	28.9&	80.5&	80.9&	67.9\\
MS2AR--3&98.6&	99.4&	89.9&	98.9&	99.2&	98.9\\
MS2AR--4&99.9&	99.9&	99.4&	99.8&	99.9&	99.9\\
MS2AR--5&84.2&	90.7&	27.9&	79.3&	80.1&	68.1\\
MS2AR--6&99.8&	99.9&	99.4&	99.8&	99.8&	99.8\\
MS2AR--7&100.0&100.0&100.0&100.0&100.0&100.0\\				
MS2AR--8&100.0&100.0&100.0&100.0&100.0&100.0\\					
MS3--1&9.6&	8.6&	3.0&	9.5&	9.6&	10.7\\
MS3--2&39.5&	34.7&	60.9&	59.0&	61.0&	62.8\\
MS3--3&83.5&	83.9&	90.3&	91.1&	90.8&      89.8\\
MS3--4&95.3&	95.2&	95.8&	95.5&	95.0&	95.6\\
MS3--5&43.9&	39.6&	60.6&	59.4&	58.9&	60.9\\
MS3--6&95.2&	95.1&	96.1&	95.8&	95.8&	95.8\\
MS3--7&97.1&	97.1&	97.1&	96.1&	96.1&	97.1\\
MS3--8&98.7&	98.7&	98.7&	97.9&	97.9&	98.7\\
MS3AR--1&12.2&	10.9&	3.3&	13.4&	12.7&	15.5\\
MS3AR--2&15.7&	12.6&	17.8&	20.4&	21.3&	25.3\\
MS3AR--3&53.9&	51.0&	67.4&	66.5&	67.7&	63.6\\
MS3AR--4&77.4&	78.3&	86.9&	88.8&	87.9&	84.2\\
MS3AR--5&18.5&	15.8&	18.1&	23.0&	23.2&	25.7\\
MS3AR--6&79.1&	78.7&	86.0&	88.6&	88.5&	84.7\\
MS3AR--7&95.4&	94.9&	95.7&	95.4&	95.4&	95.5\\
MS3AR--8&97.6&	97.6&	97.6&	96.6&	96.4&	97.5\\
\hline
\end{tabular}}
\end{center}
\begin{tablenotes}[flushleft]
\footnotesize 
\item The AR(1) coefficient, when present, is equal to 0.7 in all DGP. The transition probabilities in the MS(2) and MS(2)--AR(1) DGPs are $p_{11}=0.9$ and $p_{22}=0.8$; in the MS(3) and MS(3)--AR(1) DGPs are  $p_{11}=0.9$, $p_{12}=0.07$, $p_{22}=0.8$,  $p_{21}=0.15$, $p_{33}=0.7$, $p_{32}=0.2$. The table shows the percentage of cases, for each index, on 1000 Monte Carlo replications, where the correct number of states is identified.
\end{tablenotes}
\end{table}

\section{An Empirical Application: Identification of the Number of Regimes in U.S: Business Cycle}\label{sec:appl}
The seminal papers that sanctioned the success of MS models concern the analysis of business cycle \citep{Hamilton89,ham90}. The baseline models adopted may vary: \cite{Hamilton89} estimates an AR(4) model of quarterly GDP; \cite{Kim94} uses the state--space representation of the \cite{Lam90} model, which extends the previous \cite{Hamilton89} model with a general autoregressive component; \cite{ham_cha05} adopt a simple model as the one shown in equation (\ref{ms_2}). The common element is the number of states, set a priori equal to 2, because the aim is to interpret them as the two phases of the business cycle (expansion and contraction  respectively). 

We consider the quarterly  annualized U.S. GDP growth rate\footnote{Data are taken from the Federal Reserve Economic Data database.} (in percentage terms) from 1947:II to 2020:I, resulting in 292 observations.\footnote{We consider 2020:I as the last observation to avoid the outlier in correspondence with the COVID pandemic, which affects dramatically the stability of the estimates (as underlined by Hamilton in his Econbrowser blog: \emph{https://econbrowser.com/recession-index}).} Using fuzzy clustering, all six validation indices identify the number of states $k=3$. 

The presence of states (groups) in GDP growth rates is well established in the literature, where, as mentioned, 2-state models are generally adopted. In our opinion, and always following the parallelism with the clustering approach, it is a good practice to check for clusters in the dataset. For this purpose we adopt the homogeneity test procedure proposed by  \cite{hennig2015} to check if data are grouped or not. The procedure consists in simulating the data from a homogeneous (clusterless) DGP with parameters derived from the observed data, and calculating the cluster validation index on each of them, obtaining the empirical distribution of the index under the null hypothesis of homogeneity. The comparison of the validation index calculated on real data with the critical value derived from the simulated distribution will provide evidence of homogeneity or the presence of clusters.

We simulate 2000 series from a Normal distribution with parameters equal to the sample mean and variance of the GDP series, obtaining the p--values of the best indices (which identify 3 states) shown in Table \ref{tab:hom}. In all cases we reject the null of homogeneity (XB at 10\% significance), supporting the idea of the presence of states in the GDP growth rate series.

\begin{table}[H]
	\begin{center}
		\caption{Best values of six clustering validation indices and p--values under the null hypothesis of homogeneity}\label{tab:hom}
		\footnotesize{\begin{tabular}{ lrrrrrr } 
				\hline
				
				&PC&PE&MPC&	ASW&	ASWF&XB\\
				best value	&	0.924&	0.136&	0.886&	0.748&	0.796&	0.103\\
				p-value&	0.002&	0.001&	0.002	&0.002&	0.002&	0.078\\
				\hline
		\end{tabular}}
	\end{center}
\end{table}

 We estimate an MS model as (\ref{ms_3}) and compare it with the model proposed by \cite{ham_cha05}, corresponding to (\ref{ms_2}).  Estimation results are shown in Table \ref{tab:emp}.

We note that, in the case of 2 states, the expansion periods show an average growth rate of 3.85\% and  -1.67\% in the contraction periods. Persistence in the expansion periods is longer than in contraction periods, as shown by the probability of staying in the same state ($p_{11}$ and $p_{22}$); in this case the average duration of  state $i$ is given by $1/(1-p_{ii})$ \citep[see][]{Hamilton94}, so  the average duration of the expansion period is 20 quarters and 3.1 quarters for contraction periods. The first state of the MS(3) case shows  a very large mean (8.03), while the other 2 states are more in line with the two states of the MS(2) model. Their interpretation is immediate: state 2 is the expansion phase, state 3 the contraction phase and state 1 captures the highest peaks of the series, so that it can be interpreted as a boom period.  The average duration of the boom is 3.2 quarters, while 11.1 quarters for the expansion phase and 2.9 quarters for the contraction phase. The decrease of the variance parameter in MS(3) compared to MS(2) is clearly due to the presence of more homogeneous growth rates within each state (in practice the largest peaks are not included in the expansion state as in MS(2) model). AIC and BIC confirm that MS(3) model shows a superior performance.

\begin{table}[t]
	\begin{center}
		\caption{Parameter estimates with robust standard errors in parentheses for growth rates series of U.S. quarterly  GDP (1947:II-2000:I)}\label{tab:emp}
		\footnotesize{\begin{tabular}{cccccccccccc } 
				\hline
				$\mu_1$&$\mu_2$&$\mu_3$& $p_{11}$& $p_{12}$&$p_{21}$&$p_{22}$&$p_{32}$&$p_{33}$  & $\sigma$&AIC&BIC\\ \hline
				\multicolumn{12}{c}{MS(2)}\\
			3.85&	-1.67& & 	0.95&&&	0.68&&&3.19&5.43&5.49 \\
			(0.34)&	(1.91)&&	(0.03)&&&	(0.13)&&&	(0.19)&& 
			\\
				\hline
				\multicolumn{12}{c}{MS(3)}\\
			8.03&	2.91&-2.40 & 	0.69&0.31&0.03&0.91&0.11&0.66&2.50&5.32&5.45 \\
			(0.73)&	(0.44)&(4.00)&(0.10)&(0.10)&(0.08)&(0.08)&(0.56)&(0.42)&(0.47)&& 
			\\
				\hline

		\end{tabular}}
	\end{center}
\footnotesize{\textbf{Note}: The transition probabilities $p_{12}$ and $p_{21}$ in the 2--state cases are obtained as $p_{ij}=1-p_{ii}$ ($i,j=1,2$; $i\neq j$). The transition probabilities $p_{13}$, $p_{23}$ and $p_{31}$ are obtained as $1-\sum_{j\neq i} p_{ij}$.}
\end{table}

\begin{figure}[H]
	\caption{U.S. GDP growth rate (black line) and smoothed inference (dots) derived from  MS(2) (top panel) and MS(3) (bottom panel) models. Sample: 1947:II 2020:I.}
	\centering
	\begin{tabular}{c}
		\includegraphics[width=0.8\linewidth,height=0.5\textwidth]{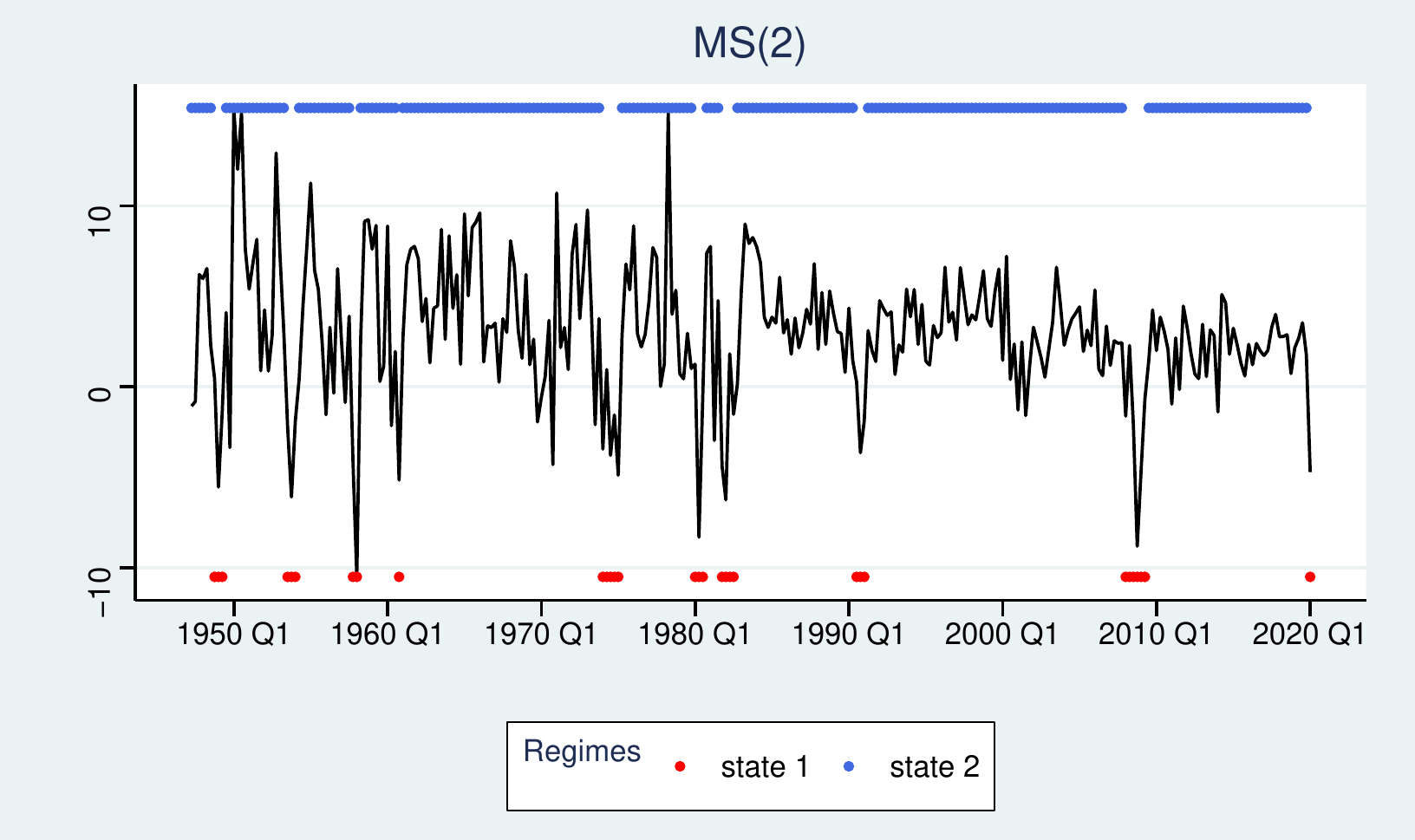}\\
		\includegraphics[width=0.8\linewidth,height=0.5\textwidth]{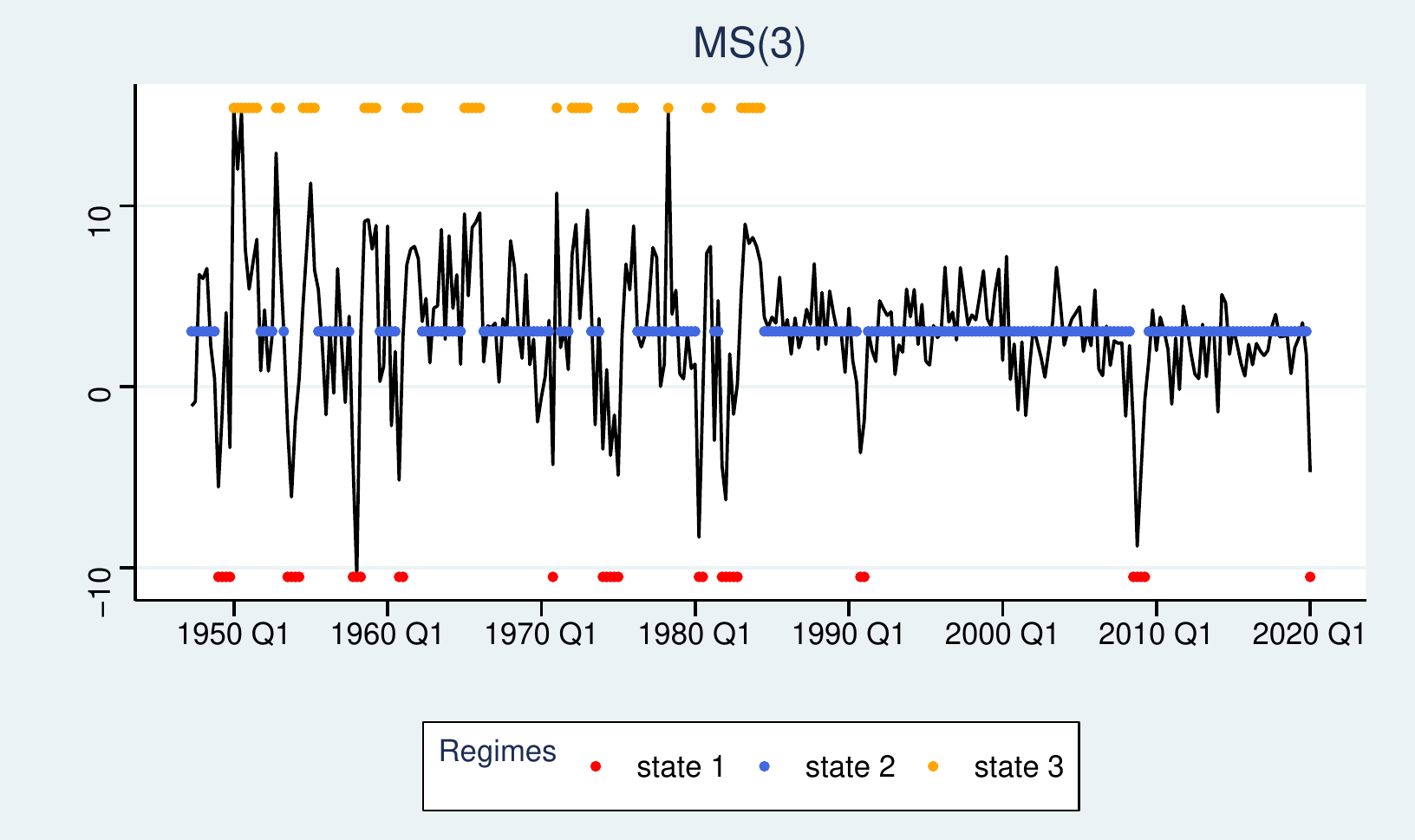} 
	\end{tabular} \label{fig:smooth}
\end{figure}

Previous comments are supported by Figure~\ref{fig:smooth}, where we show the inference on the state derived form the smoothed probabilities, both for MS(2) (upper panel) and MS(3) (lower panel).   It is clear that state 2 in MS(2) is equivalent to state 3 in MS(3) and corresponds to the contraction periods, while state (1)  in MS(2) is decomposed in state 1 (corresponding to the highest peaks) and state 2 (moderate positive values) in MS(3).   

Finally, we calculate the Rand index to compare the inference on the state derived from MS(3) model and fuzzy clustering. The value is 0.74, which is consistent with the MS3--1 case illustrated in Table \ref{tab:rand}. Of course the great variability of the data affects this result, in particular for the presence of abrupt changes  in the dynamics of the time series. In fact, smoothing the original series with moving averages, the Rand index grows rapidly: a 3--term moving average implies a Rand index of 0.81, while a 5--term moving average shows a Rand index  equal to 0.90, confirming the evidence for the presence of 3 states. 

\section{Final Remarks}\label{sec:concl}
Detecting the number of states $k$ is a crucial task in estimating MS models. This problem is often circumvented, by fixing $k$ a priori, due to the impossibility of applying the classical statistical tests for the problem of nuisance parameters present only under the alternative hypothesis.  In this work we underline the similarity between the idea of inference on the state, obtained from MS models, and fuzzy clustering. After verifying the similarity of the grouping with the two procedure through Monte Carlo experiments, on the same simulated series we verified the ability of the main clustering validation indices to detect the correct number of states. Our experiments show a good performance, in particular of two indices (PE and ASW), also when the dynamics of the time series is affected by autocorrelation; two other indices (MPC and XB) show some puzzling behavior. The performance of the indices increases when the differences between the means within states is greater and the variability is lower.

The results seem to support the possibility of adopting  clustering validation indices as a tool to identify, before the estimation step, the number of states of MS models in a non--parametric way. The advantage of this procedure is even more evident if we consider that the calculation of these indices is implemented in the main statistical packages.

We think that when different validation indices identify the same number of states, there is strong support for this result. In the last section we tried to apply this procedure to the US GDP growth rate series, identifying 3 states with all six indices, against the usual practice of setting (a priori) $k=2$ for this variable.

As future research it might be interesting to consider other distance measures, such as that of Mahalanobis, or medoid-based approaches as opposed to the centroid-based ones proposed here. It would thus be possible to compare the different fuzzy clustering algorithms to evaluate which one works better.

From another perspective, it might be interesting to compare fuzzy clustering with the regime inference obtained from the MS model with fuzzy regimes, recently proposed by \cite{go18}, where the separation between regimes is not clear-cut, but with the possibility of overlapping states.

\renewcommand{\baselinestretch}{1.1} \selectfont
\bibliographystyle{chicago}
\bibliography{biblio}
\end{document}